\documentclass{aa}  
\usepackage{amsmath} 

\usepackage{graphicx}
\usepackage{txfonts}
\begin{document}

   \title{The \ion{Ca}{II} Infrared Triplet's performance as an activity indicator compared to \ion{Ca}{II}~H~\&~K}

   \subtitle{Empirical relations to convert \ion{Ca}{II} Infrared Triplet measurements to common activity indices.}

   \author{J. Martin
          \inst{1}
          \and
          B. Fuhrmeister\inst{1}
          \and
          M. Mittag\inst{1}
          \and
          T. O. B. Schmidt\inst{1,2}
          \and
          A. Hempelmann\inst{1}
          \and
          J. N. Gonz\'alez-P\'erez\inst{1}
          \and
          J. H. M. M. Schmitt\inst{1}
          }

   \institute{Hamburger Sternwarte, Universität Hamburg, 21029 Hamburg,
              Germany
              \and LESIA, Observatoire de Paris, CNRS, PSL Research University, Sorbonne Universit\'{e}s, UPMC Univ. Paris 06, Univ. Paris Diderot, Sorbonne Paris Cit\'{e}, 5 place Jules Janssen, 92190, Meudon, France\\
              \email{johannes-martin@hs.uni-hamburg.de}
              }

   \date{Received 21 December 2016 / Accepted 19 May 2017}
 
  \abstract
   {}
   {A large number of Calcium Infrared Triplet (IRT) spectra are expected from the GAIA- and CARMENES missions. Conversion of these spectra into known activity indicators will allow analysis of their temporal evolution to a better degree.
   We set out to find such a conversion formula and to determine its robustness.}
   {We have compared 2274 \ion{Ca}{II}~IRT spectra of active main-sequence F to K stars taken by the TIGRE telescope with those of inactive stars of the same spectral type.
    After normalizing and applying rotational broadening, we subtracted the comparison spectra to find the chromospheric excess flux caused by activity. We obtained the total excess flux, and
    compared it to established activity indices derived from the \ion{Ca}{II} H \& K lines, the spectra of which were obtained simultaneously to the infrared spectra.
    }
   {The excess flux in the \ion{Ca}{II}~IRT is found to correlate well with $R_\mathrm{HK}'$ and $R_\mathrm{HK}^{+}$, as well as $S_\mathrm{MWO}$, if the $B-V$-dependency is taken into account. 
    We find an empirical conversion formula to calculate the corresponding value of one activity indicator from the measurement of another, by
    comparing groups of datapoints of stars with similar $B-V$.   
   }
   {}


   \maketitle
%

\section{Introduction}

Cool stars with outer convective envelopes ubiquitously show signatures of magnetic activity. Such activity manifests itself in a plethora of observable signatures, such as spots, chromospheric emission lines, emission at X-ray and XUV wavelengths, and many others.  On quite a few stars these activity phenomena are more pronounced than what we observe in the Sun, and it is therefore useful to perform activity studies on other, more active stars, both to learn about the underlying physical processes, but also to learn more about the Sun.\\
One of the best-known measures of activity is the so-called Mount-Wilson S-index $S_\mathrm{MWO}$, defined as the ratio of the flux in the center of the \ion{Ca}{II} H \& K lines, where activity results in a 
sometimes very large excess emission, relative to the flux in the continuum on either side of the lines. 
As we have access to a large number of such observations dating back many years, this S-index is well-suited for long-term activity studies \citep{Duncan91}. It has, in fact, been used to determine periods for activity cycles and/or rotation in cool stars \citep{Baliunas95}. \\
Since the photosphere also contributes in the center of the \ion{Ca}{II} H \& K lines, the S-index characterizes not only chromospheric activity, and it becomes difficult to compare stars with different
effective temperatures, where these photospheric contributions will vary. To overcome these shortcomings, \citet{Linsky79} introduced the so-called $R_\mathrm{HK}'$-index.
The photospheric flux is first subtracted from the flux measured in the \ion{Ca}{II} H \& K lines, and the remainder subsequently normalized by dividing by $\sigma T_\mathrm{eff}^4$. This correction allows a direct comparison of stars of various stellar types, which have different photospheric fluxes.
Given $T_\mathrm{eff}$, it is possible to convert the measured values of $S_\mathrm{MWO}$ into $R_\mathrm{HK}'$ \citep{Rutten84,Linsky79}, hence the large amount of archival data for $S_\mathrm{MWO}$ can directly be used to compare in the $R_\mathrm{HK}'$-scale.\\
Both the GAIA mission \citep{Prusti2012} and CARMENES \citep{Carmenes2014} are expected to provide very large numbers of spectra that can be used for activity studies of stars. 
The Radial Velocity Spectrometer (RVS) onboard GAIA has a resolution of about 11\,500 with a wavelength coverage of 8470-8740\,\AA, and CARMENES covers the region between 5\,500--17\,000\,\AA~with a resolution of 82\,000. The GAIA RVS is expected to yield spectra down to a magnitude of about 17, which corresponds to 15-16 \% of the GAIA catalog of presently 1\,142\,679\,769 entries \citep{GAIA_DR1}. CARMENES will yield time series
of selected M dwarfs and the total number of spectra in the first three years will be approximately 15\,000.\\
In both cases, the \ion{Ca}{II} H \& K lines at 3933.7\,\AA~and 3968.5\,\AA~are not covered, and thus no data to enhance temporal studies of activity can be combined with the existing $S_\mathrm{MWO}$ data. 
However, spectra obtained with either of these instruments cover the Calcium Infrared Triplet (IRT), three lines centered at 8498\,\AA, 8542\,\AA~and 8662\,\AA. Like the \ion{Ca}{II}~H~\&~K lines, the IRT are \ion{Ca}{II} lines, which have been reported to be sensitive to activity as well \citep{MartinzesArnaiz11}. In contrast to the \ion{Ca}{II} H \& K lines, these IRT lines usually only show a smaller fill-in due to activity, rather than stronger fill-in up to a clear emission core as the \ion{Ca}{II} H \& K lines do. 
While the rather simple indicator of the central depression in the IRT lines correlates well with $S_\mathrm{MWO}$ \citep{Chmielewski2000}, it is difficult to disentangle the effects from activity on the central depression from those of rotational broadening, which is also known to correlate well with activity \citep{Andretta05}. \citet{Busa07} presented a new indicator $\Delta W_\mathrm{IRT}$, based on the excess flux from the chromosphere, obtained by subtracting a model of the photosphere. Their new indicators turned out to correlate well with $R_\mathrm{HK}'$, and they were able to obtain a conversion formula from this index to $R_\mathrm{HK}'$ by analyzing the spectra of 42 stars of type F5 to K3.\\
In this paper, we have adopted a similar approach, but have subtracted the observed spectra of an inactive object with similar stellar parameters. We performed this comparison not only
for the calcium lines, but also for H$\alpha$, and we also fit a Gaussian to the obtained excess flux to test the feasibility of using such a fit as an activity indicator.
We used more than two thousand observations obtained with the TIGRE telescope, which simultaneously records the \ion{Ca}{ii} H \& K lines, the \ion{Ca}{ii}~IRT and H$\alpha$ (see Sect. \ref{sec:tigre}). This means that there is no scatter 
from temporal variation, which can be a rather significant introduced error \citep{Baliunas95}, and yet we are also able to more accurately quantify the expected derivation simply from inherent differences in the two activity indices.\\
The plan of our paper is as follows: First, we describe the TIGRE telescope and give an overview of the objects and observations used in this paper. We then describe the method used to determine the excess flux, and show the results for the measured line flux for inactive objects. Then, we show the observed correlations for the excess flux in the lines to other indicators. Finally, we give relations to convert the measured values to other indicators.

\section{Observations}

\subsection{The TIGRE telescope}
\label{sec:tigre}

The Telescopio Internacional de Guanajuato Rob\'otico Espectrosc\'opico (TIGRE) is operated by a collaboration between the Hamburger Sternwarte, the University of Guanajuato and the University of Li\`ege. TIGRE is a 1.2~m telescope stationed
at the La Luz Observatory in central Mexico near Guanajuato 
at a height of about 2400~m. 
Equipped with the refurbished Heidelberg Extended Range Optical Spectrograph (HEROS), 
TIGRE takes spectra with a resolution of $\sim$20\,000, covering a wavelength range
of $\sim$3800-8800\,\AA, with only a small gap of about 130\,\AA~centered at 5765\,\AA. This wide wavelength coverage allows to obtain measurements of the \ion{Ca}{II}~IRT 
simultaneously with those taken of the \ion{Ca}{II} H \& K lines.
TIGRE can be operated both manually and fully automatically, 
including the selection of the observation time for each object, based on factors such as weather, position, visibility in other nights, and the assigned priority. More detailed information about TIGRE can be found in \citet{Schmitt2014}.\\
After every night the system automatically reduces the data, running a modified version of the REDUCE package \citep{Piskunov2002}, as described in \citet{Mittag2010}. This reduction pipeline includes flatfielding and the wavelength calibration.
Moreover, TIGRE determines its own
S-index, defined almost identically to the original Mount-Wilson S-Index $S_\mathrm{MWO}$ \citep{Vaughan78,Duncan91},
\begin{equation}
 S_\mathrm{MWO} = \frac{N_H + N_K}{N_V + N_R} \alpha ,
 \label{eq:sindexdef}
\end{equation}
where $N_H$ and $N_K$ are the countrates in a bandpass with a FWHM of 1.09\,\AA~in the center of the \ion{Ca}{II} H \& K line, respectively. In the original definition, this bandpass is triangular, whereas the TIGRE S-Index uses a rectangular bandpass. $N_V$ and $N_R$ are the countrates in 20\,\AA-wide continuum bands outside the lines, centered at 3901.07\,\AA~and 4001.07\,\AA. The factor $\alpha$ ensures that countrates measured by different instruments are in agreement. 
The TIGRE S-Index can be converted to the $S_\mathrm{MWO}$-scale \citep{Mittag2016}. 

\begin{figure}
\resizebox{\hsize}{!}{\includegraphics{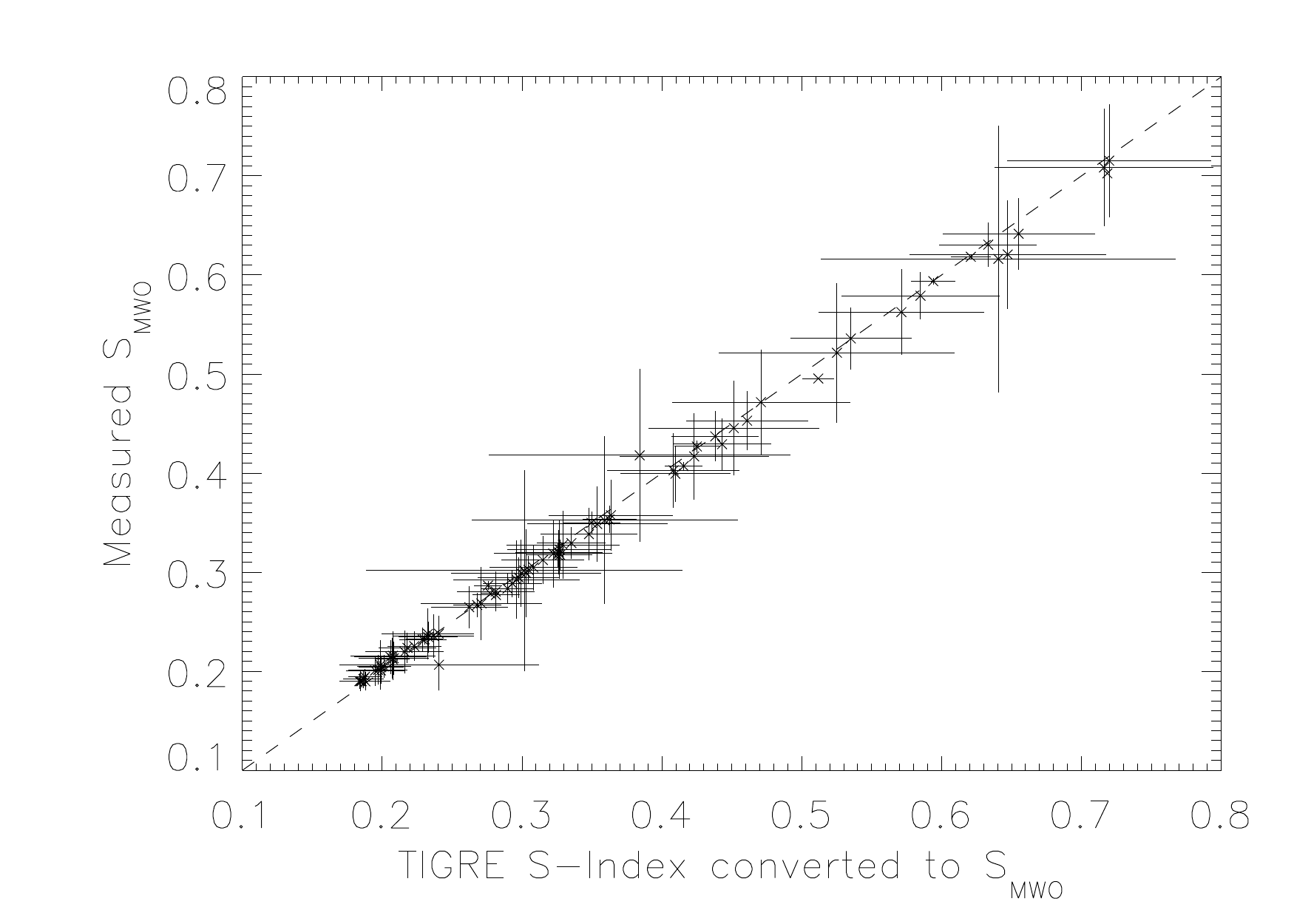}}
\caption{Comparison of our measured value for $S_\mathrm{MWO}$ with the converted TIGRE S-Index.}
\label{fig:mystosmwoconv}
\end{figure}

In this paper, we have measured the S-index ''manually'' from the spectra for every observation using the same 
bandpasses as given in the original definition, including the triangular shape in the center of the lines. To determine $\alpha$ correctly for our values, we compare our values to the corresponding TIGRE S-index values converted to $S_\mathrm{MWO}$. 
As shown in Fig.~\ref{fig:mystosmwoconv}, there is a clear linear relation between 
the two S-indices, allowing us to transform our values to the $S_\mathrm{MWO}$-scale.
We cannot simply always use the TIGRE-determined S-index, because older versions of the pipeline did not calculate that value. To ensure that we can also use these spectra, but do not introduce systematic errors due to a different approach in calculating the S-index, we have always calculated it manually according to the original definition.

\subsection{Overview of data used}

In this paper, a total of 2807 individual observations of 102 stars were analyzed. Some of these observations were not suitable for the excess flux determination (see Sect. \ref{sec:outliers}), so that only {2274} observations of 82 stars were eventually used.
The stars with observations used here were not originally selected for this paper, but rather for other science purposes.
The largest part of the data was originally taken to determine stellar rotation periods of solar-like stars \citep{Hempelmann16}. Here, we only look at data from main-sequence
stars with $B-V$-colors ranging from 0.4 to 1.2, corresponding to F, G and K stars. The earliest data points are from April 15, 2013, ranging up to the latest from May 4, 2016. We have excluded data points obtained between December 6th, 2014 and May 15th, 2015, since there was a different camera for the red channel in use at that time. 
Because these objects were observed for different projects, the signal-to-noise-ratio (S/N) and exposure time are not constant in our sample. 
We only analyzed observations with an average S/N of at least 20, because 
otherwise their noise level introduces large errors in our sample.
Finally, telluric line correction was done using Molecfit \citep{Smette2015,Kausch2015}.
We have used the stellar parameters given in \citet{Soubiran2010} whenever possible. 
Table~\ref{tab:dataoverview} shows an overview of the number of analyzed observations for each spectral type, as well as the minimum, median and maximum values for S/N and exposure time in that class. We provide a full list of all objects, with sources for values $B-V$, $v\sin{i}$, $\log{g}$ and [Fe/H] that we used, in Table~\ref{tab:objparas} and 
Table~\ref{tab:compparas} in the Appendix.
\begin{table}
\caption{Overview of the data used in this paper, categorized by spectral type.}
\label{tab:dataoverview}
\centering
\begin{tabular}{rllll}
\hline
\hline
\textbf{Type} & \textbf{\#} & \textbf{\#} & \textbf{S/N} & \textbf{Exp. time} [s]\\
 & \textbf{Obj.} & \textbf{Obs.}  & \textbf{min/med/max} & \textbf{min/med/max}\\
\hline
F & 9 & 265 & 36.1 / 60.7 / 161.3&120 / 360 / 2578\\
G & 46 & 1419 & 20.1 / 59.2 / 114.0&60 / 622 / 4767\\
K & 27 & 590 & 20.4 / 64.3 / 114.8&60 / 799 / 4846\\
\hline
\textbf{Total} & {82} & {2274} & 20.1 / 60.4 / 161.3 & 60 / 610 / 4846\\
\end{tabular}
\end{table}

\section{Method}

\subsection{Selecting comparison objects}

The changes in the \ion{Ca}{II}~IRT lines due to activity are much smaller 
than those seen in the \ion{Ca}{II} H \& K lines. 
To measure this change, we compared our observations with those from an inactive star.
The comparison star must be similar in its parameters to the active star 
in question, to ensure that the difference in the line profiles stems 
from activity rather than from differences in the photosphere. 
Whether a star is considered active or inactive is determined by its value of $R_\mathrm{HK}'$, defined by \citet{Linsky79} as:
\begin{equation}
\label{eq:rhkdef}
 R_\mathrm{HK}'  =  \frac{F_\mathrm{HK}-F_\mathrm{HK,phot}}{\sigma T_\mathrm{eff}^4},\\
\end{equation}

where $F_\mathrm{HK}$ is the flux measured in the \ion{Ca}{ii} H \& K lines and $F_\mathrm{HK,phot}$ the photospheric contribution to that flux. 
Since this index is normalized to $\sigma T_\mathrm{eff}^4$, it is only marginally dependent on $B-V$, and thus, while slightly more difficult to determine, better suited for activity studies. We have used the relation given in \citet{Mittag2013} to convert our measured values for $S_\mathrm{MWO}$ to $R_\mathrm{HK}'$. We only study stars with $\log{R_\mathrm{HK}'} \geq -4.75$, and define those 
with smaller $R_\mathrm{HK}'$ as inactive, following the definition by \citet{Henry96}. 
This threshold value is close to the lower levels of the Sun's activity. 
For each potentially active star in question, we select one inactive star as close as possible in stellar parameters and slowly rotating, so that $v\sin{i} \leq 5~\mathrm{km}\mathrm{s}^{-1}$. These criteria have been given different weights: 
A similar value for $B-V$ is given the highest priority, followed by similar values for metallicity and then gravity.
For each comparison object, the ``best'' comparison spectrum - defined as the one with the highest S/N - is selected, and visually checked to ensure that no artifacts remain, for example from uncorrected cosmics.
Every observation of the star in question is compared to that spectrum, referred to as comparison spectrum in the following.

\subsection{Errors from incorrect stellar parameters}
\label{sec:wrongstellarparas}

\begin{figure}
\resizebox{\hsize}{!}{\includegraphics{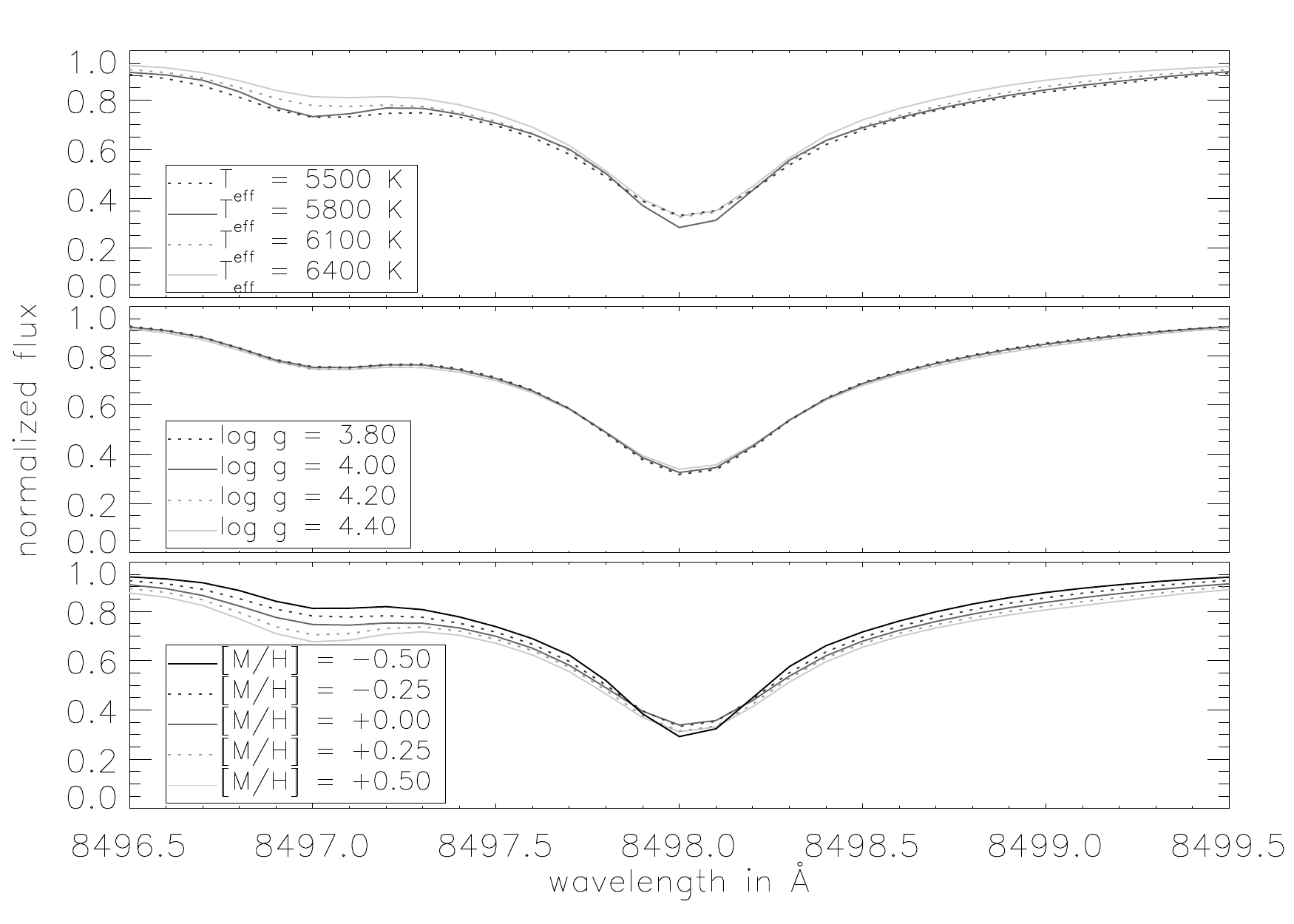}}
\caption{Effects of stellar parameters on the \ion{Ca}{II} IRT lines. Shown here are PHOENIX spectra, degraded to the resolution of TIGRE spectra, with $T_\mathrm{eff}$ varied in the top plot, $\log{g}$ in the middle, and metallicity in the bottom plot. Unless varied in that plot, stellar parameters were set to $T_\mathrm{eff}=5700\,\mathrm{K}$, $\log{g}=4.40$ and $\left[\mathrm{M/H}\right]=0.0$.}
\label{fig:phoenix_irtparams}
\end{figure}

The stellar parameters are not always well-determined, and sometimes even a rather large range of possible values is given in the literature. The \ion{Ca}{II} IRT line profiles, especially the wings, are affected quite strongly by changes in metallicity as analyses of model spectra show \citep{Smith87,ErdelyiMendes91}. In Fig. \ref{fig:phoenix_irtparams}, we show the normalized spectra of the first \ion{Ca}{II} IRT line from several interpolated PHOENIX models \citep{Hauschildt99}, that are based on those from \citet{Husser2013}. In these spectra, \ion{H}{i}, \ion{He}{i}, \ion{He}{ii}, \ion{Ca}{i}, \ion{Ca}{ii}, \ion{Ca}{iii}, \ion{Fe}{i} and \ion{Fe}{ii} were, among others, all calculated in Non Local Thermodynamic Equilibrium (NLTE). The plotted first \ion{Ca}{ii} IRT line shows the strongest effects, and allows us to give a ``worst-case'' estimate. It is obvious that the lines are not very sensitive to gravity, but show a strong dependence on metallicity, confirming the result by \citet{Andretta05}. 
Temperature also
affects the line profile, but values for $T_\mathrm{eff}
$ tend to be determined more reliably.\\
Most of the stellar parameters we have used are taken from \citet{Soubiran2010}, where the authors have compiled the stellar parameters from the literature. The average discrepancy in metallicity for stars that have more than one set of stellar parameters available is given there as 0.08\,dex, and the discrepancy in $T_\mathrm{eff}$ as 1.3\,\%. For our rough determination on the errors introduced in the final excess flux, we ignored the low discrepancy on $T_\mathrm{eff}$, as we can confidently say from Fig. \ref{fig:phoenix_irtparams} that a deviation of $\sim10\,\mathrm{K}$ will have neglibile effects compared to those from the deviation in metallicity. We then considered the conservative case of template and active star to both have incorrectly determined metallicity, and that the real difference between the two is $\Delta \left[\mathrm{M/H}\right]=0.25\,\mathrm{dex}$. We took two model spectra with $T_\mathrm{eff}=5700\,\mathrm{K}$, $\log{g}=4.40$ and [M/H]=0.25 and [M/H]=0, respectively, and integrated 
the 
flux of 
those two spectra 
numerically, across 
a 1\,\AA-wide bandpass in the center of the first \ion{Ca}{II} IRT line. The results differ by less than 3\,\%. We therefore conclude that the error from incorrect stellar parameters will not strongly affect our results.

\subsection{Comparing active stars to inactive template stars}
\label{sec:comparing}

The \ion{Ca}{II} H \& K lines, the \ion{Ca}{II} IRT lines, and H$\alpha$ were checked individually with a procedure (written in IDL) that worked as follows: For each line, a region was defined that encompasses the line and continuum on either side. 
Both observation and comparison spectrum were then normalized in this region by finding a linear fit to the upper envelope in small regions defined as continuum.
Observation and comparison spectrum were then shifted on top of each other by cross-correlation. In this way, any potential wavelength-shift, no matter the cause, is corrected.
Afterwards, the comparison is rotationally broadened to the rotational velocity of the actual star, following the procedure described by \citet{Gray2005}, with a limb-darkening coefficient interpolated from the figure given there (their Fig. 17.6, pg. 437). 
As a local normalization, a fit was performed to match the photospheric wings.
This can be done, since only the core of the line should be affected by chromospheric activity \citep{Busa07}. Finally, we subtracted the comparison spectrum from the spectrum, and ended up with the excess flux, thought to come
from chromospheric activity. We integrated this excess curve in an 1\,\AA-wide region for the \ion{Ca}{ii} IRT lines and H$\alpha$, and a 2\,\AA-wide region for \ion{Ca}{ii}~H\,\&\,K to obtain the resulting excess flux $F_\mathrm{Exc}$. The larger bandpass for the \ion{Ca}{ii}~H\,\&\,K-lines has been purposefully selected to be larger than the expected width of $\sim$\,1\AA~according to the Wilson-Bappu-effect \citep{Wilson1957}, to ensure that all of the flux is included in these rather wide lines. 

\begin{figure}
\centering
\resizebox{\hsize}{!}{\includegraphics{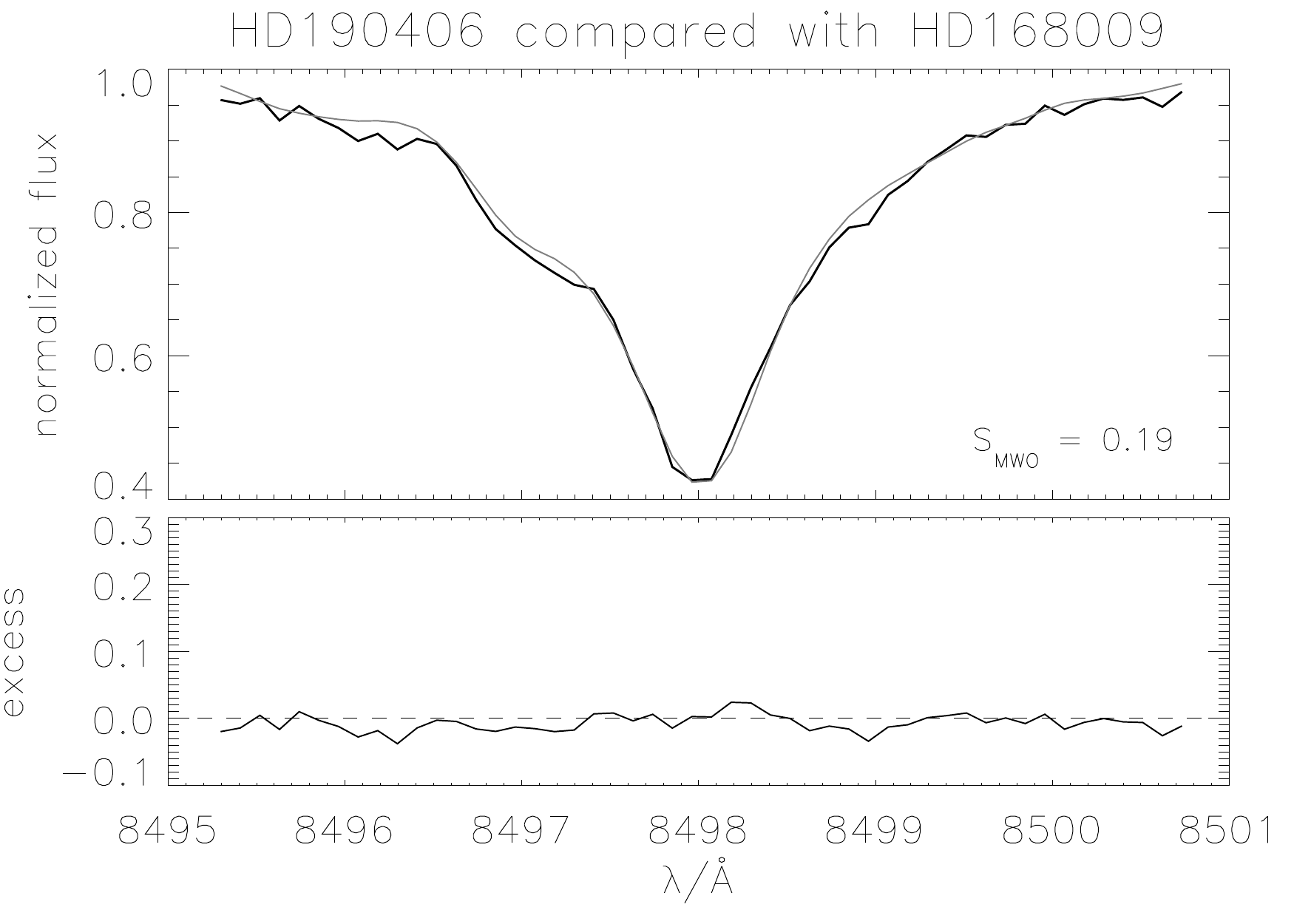}}
\resizebox{\hsize}{!}{\includegraphics{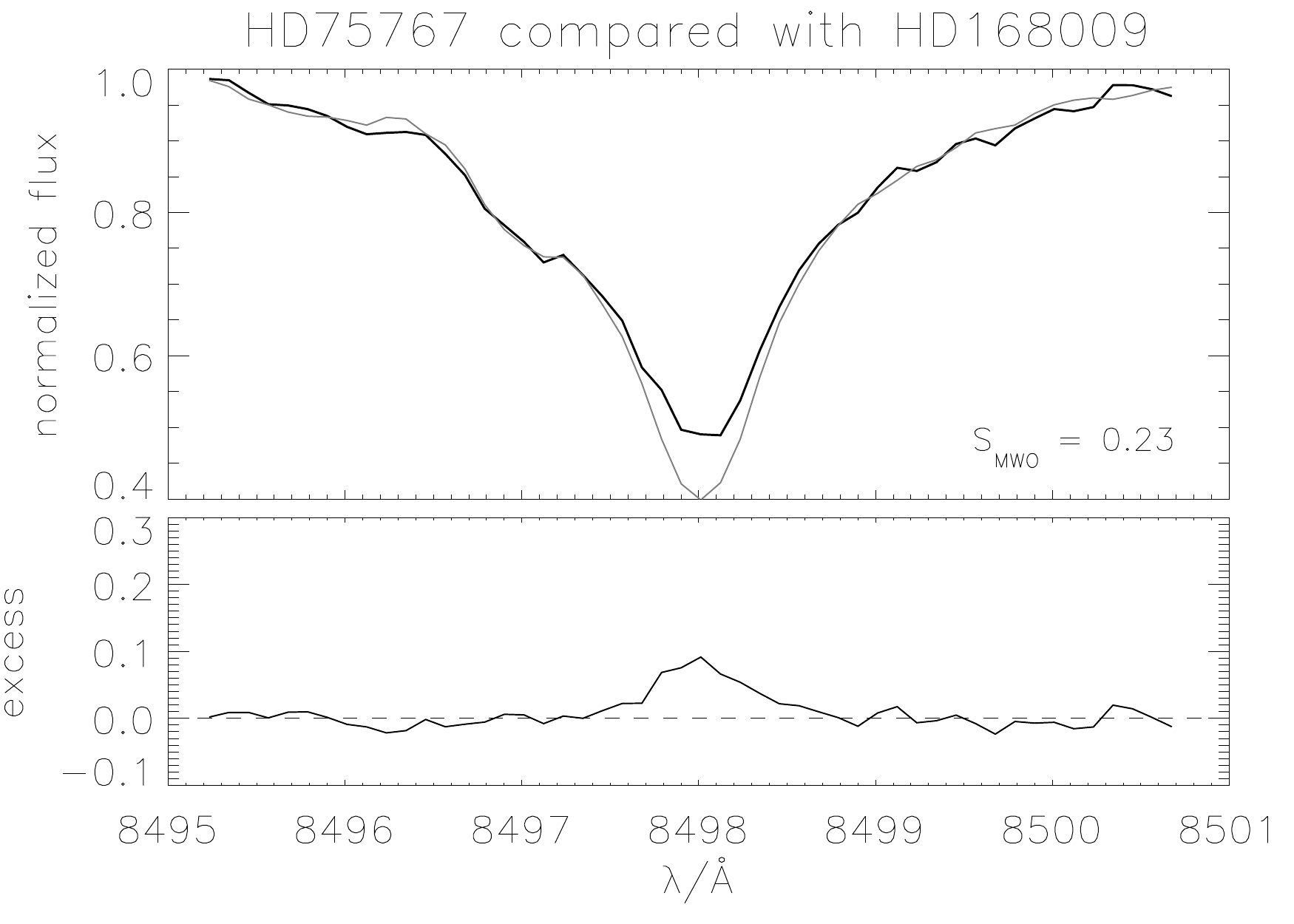}}
\resizebox{\hsize}{!}{\includegraphics{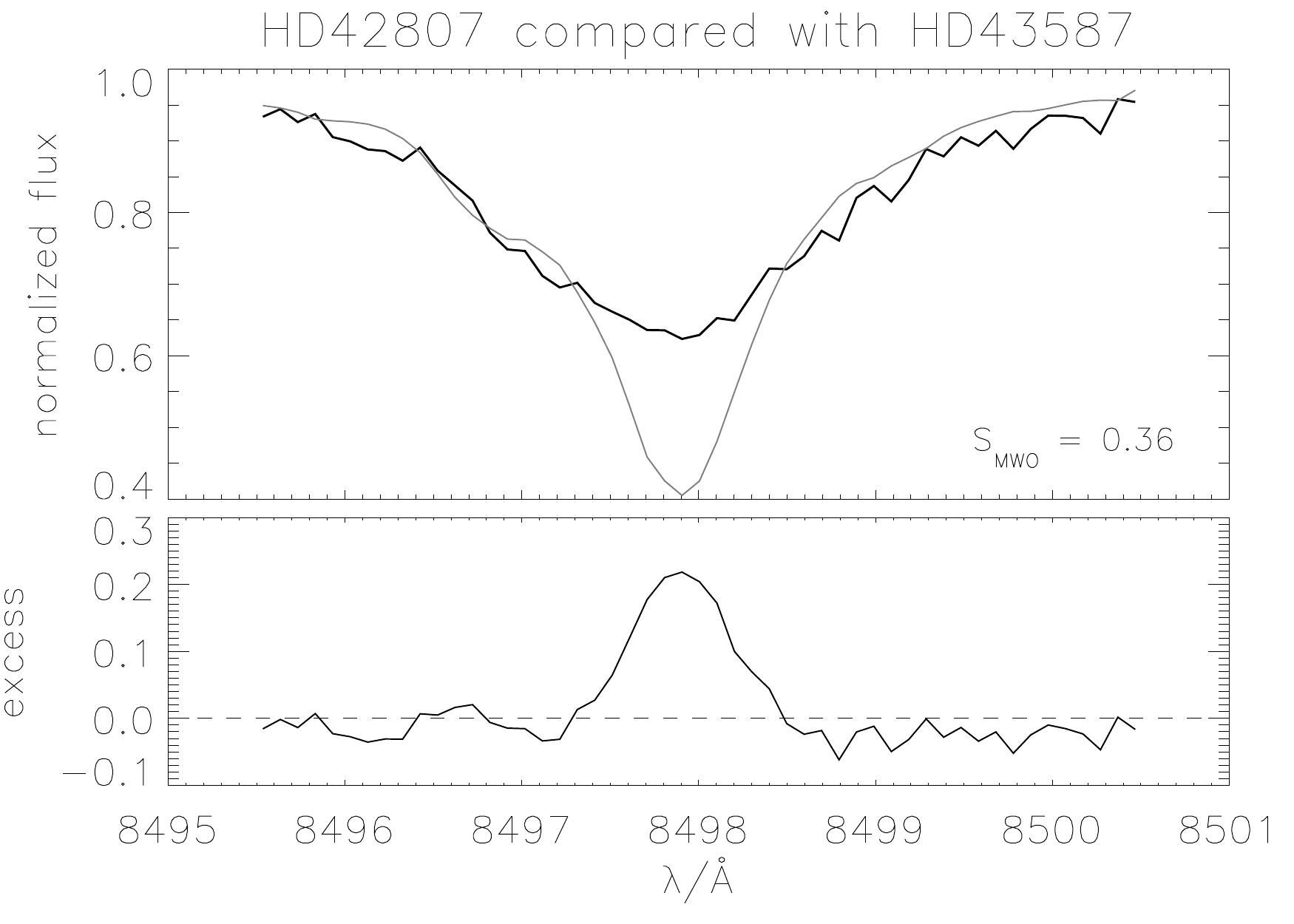}}
\caption{Comparison of different stars' spectra (black) with the spectra of inactive stars (gray). For the lowest-activity stars (\textit{Top}), no excess flux can be seen, whereas for higher activity, the observed fill-in increases with activity. The determined excess flux in the first \ion{Ca}{II} line in these three cases are $0.1\cdot10^5$\,erg\,s$^{-1}$\,cm$^{-2}$, $2.0\cdot10^5$\,erg\,s$^{-1}$\,cm$^{-2}$ and $4.4\cdot10^5$\,erg\,s$^{-1}$\,cm$^{-2}$.}
\label{fig:compexamples}
\end{figure}

\begin{figure}
\resizebox{\hsize}{!}{\includegraphics{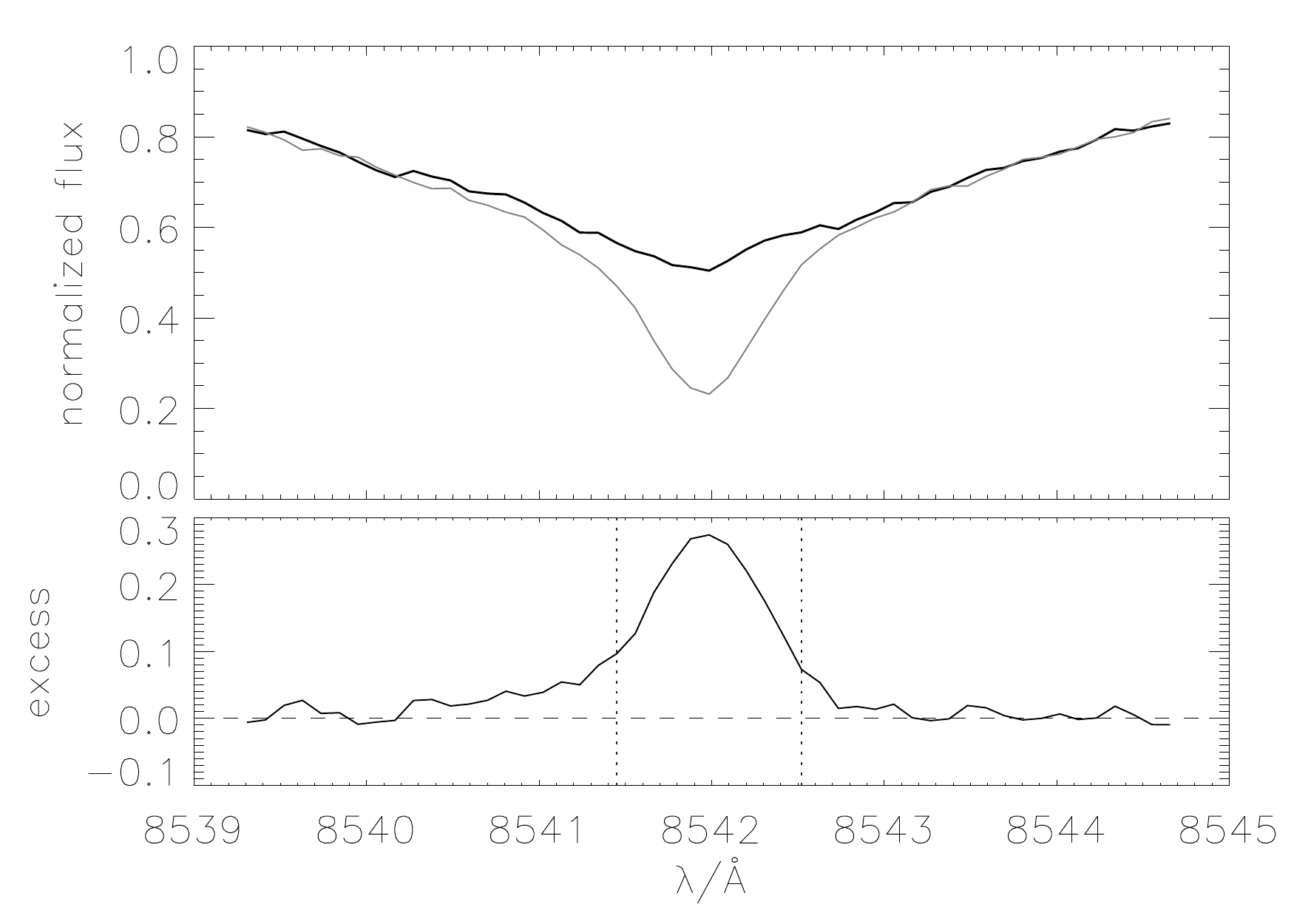}}
\resizebox{\hsize}{!}{\includegraphics{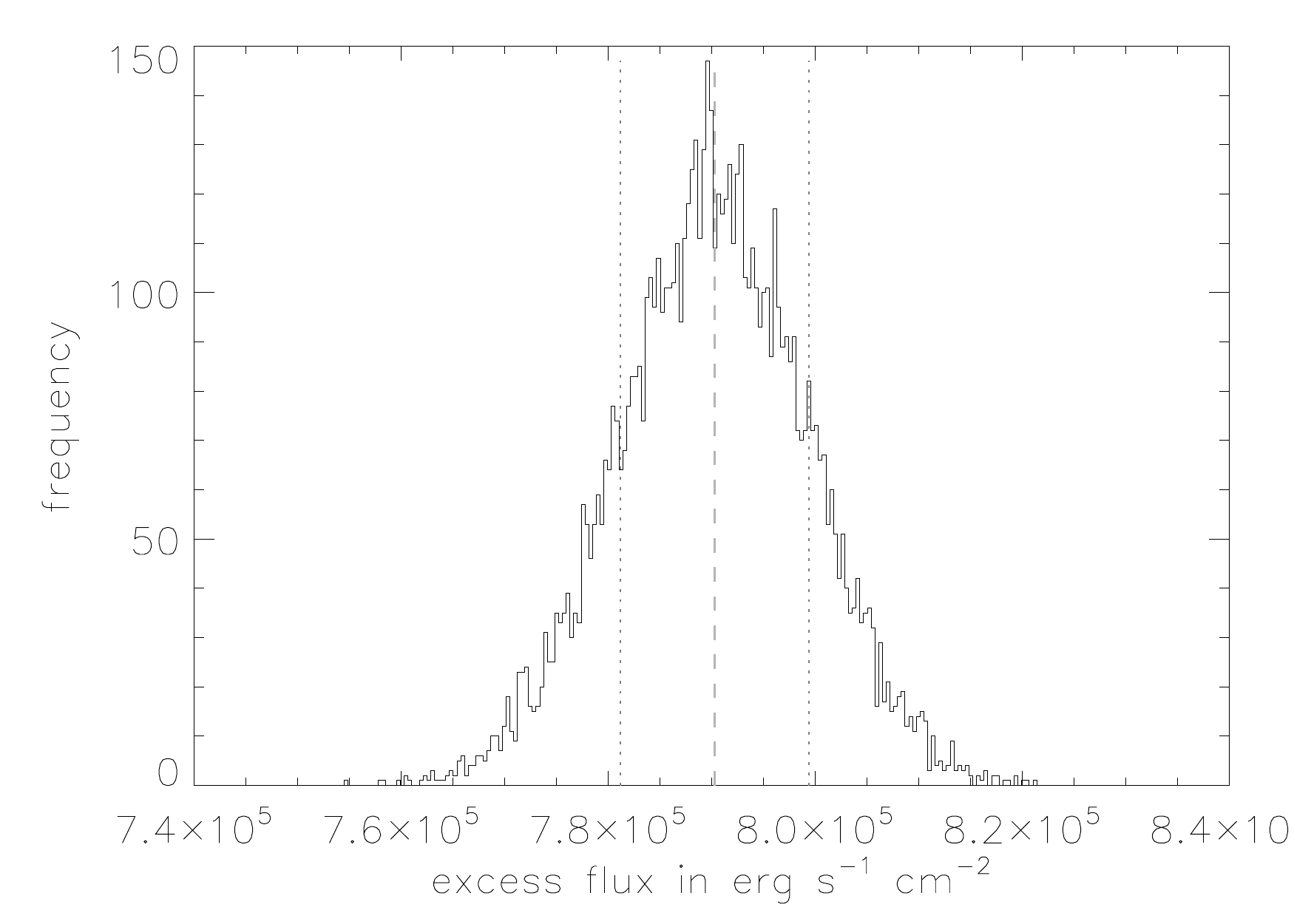}}
\caption{Determination of the excess flux measured for HD152391 in the second \ion{Ca}{II}~IRT line at 8542\AA~and resulting distribution of the integrated excess flux after performing the Monte-Carlo-iterations. \textit{Top:} Comparison of the observation (black) with the rotationally broadened comparison spectrum (gray). The resulting excess flux, shifted to find an agreement in the wings, is shown in the lower plot. The dotted vertical lines show the 1\AA-region used for integrating. \textit{Bottom:} Histogram resulting from performing the excess flux determination 150 times, varying rotation and other parameters within their errors. The gray dashed line shows the average excess flux, the gray dotted lines the found $1\,\sigma$ values. Here, we found an excess flux of $(7.9\pm0.09)\cdot10^5$\,erg\,s$^{-1}$\,cm$^{-2}$.}
\label{fig:excessdistribution}
\end{figure}

In Fig.~\ref{fig:compexamples} we show the result of such a comparison for 
three objects with different levels of activity, i.e., for a star with low activity ($S_\mathrm{MWO}$ = 0.19), medium activity ($S_\mathrm{MWO}$ = 0.23) and high
activity ($S_\mathrm{MWO}$ = 0.36).  Figure~\ref{fig:compexamples} also shows
that the observed excess flux in the lines is increased for the more 
active objects. The excess fluxes shown here correspond to $0.1\cdot10^5$\,erg\,s$^{-1}$\,cm$^{-2}$, $2.0\cdot10^5$\,erg\,s$^{-1}$\,cm$^{-2}$ and $4.4\cdot10^5$\,erg\,s$^{-1}$\,cm$^{-2}$ respectively.
In order to assess the error of the excess flux, we use Gaussian error propagation where possible, for example propagation of the errors on the normalization fit, or a
Monte Carlo-approach, for example by broadening the line 150 times with the 
values for $v\sin{i}$ varying within its error; in those cases where no error 
is given, we have assumed a 10\,\%~error.
The resulting distribution of the values for the integrated excess flux are 
Gaussian in almost every case, so that we interpret the resulting error 
as a $1\,\sigma$-error; see Sect. \ref{sec:outliers} for a description 
of objects for which the distribution is not Gaussian. 
A typical example for our procedure is shown in Fig.~\ref{fig:excessdistribution}.
Since the spectra are normalized to unity, this implies that this excess flux 
is given in units of the continuum flux.  To convert this to a stellar surface flux in units of erg\,cm$^{-2}$\,s$^{-1}$\,\AA$^{-1}$, we use the relation 
from \citet{Hall96} for the continuum flux at different wavelength points.
We also fit Gaussians or Lorentzians to the resulting excess flux distribution using
the MPFIT routine \citep{Markwardt2009}. These fit parameters can in principle
also be used as activity indicators (see Sect.~\ref{sec:fittedexcess}). 
The equivalent width has also been determined, but gives less reliable results 
than the integrated flux.
We present a formula to determine the flux of inactive objects 
in Sect.~\ref{sec:firtinactive}, which can be subtracted from a measured flux to estimate the excess flux.\\

In some cases, we obtain a negative value for the excess flux, implying that the comparison star was more active than the star under investigation. Indeed, this only occured
for observations that also feature a low value for established activity indicators, as well as a higher noise level. Our procedure is not well suited for these objects, because the 
change in the line that stems from activity is smaller than the errors introduced 
from the line profile differences as the stellar parameters of the two objects do not match exactly. A future study that compares the spectra to models will hopefully resolve this.\\

As mentioned earlier, \citet{Busa07} introduced a new activity index $\Delta W_\mathrm{IRT}$, which is obtained in a similar fashion. Two objects studied by \citet{Busa07} are also in our sample: \object{HD\,25998} and \object{HD\,82443}. A slight change in our calculation allows us to also obtain this parameter from our data. For both objects and all three lines, the values agree to within $1\,\sigma$.

\subsubsection{Models or inactive stars as comparison?}

In our approach we have chosen to make use of the large sample of available stellar TIGRE spectra, and to compare the spectra of active stars to the spectra of inactive stars. This approach requires no further assumptions on the formation of the spectrum. One advantage of this approach compared to the one of subtracting a model spectrum, as is the case in \citet{Busa07}, is that we remove the basal flux level as well, leaving only the ``true'' activity related excess flux.
An additional advantage is that we avoid errors due to incorrect parameters in the line list, or incomplete or otherwise erroneous line profiles. This error is hard to quantify and likely to be systematic in nature. On the other hand, the observed spectra of inactive stars will have a certain degree of noise in them, which introduces some scatter as well. However, across a large sample, these errors are statistically distributed, and average out.\\
Since the comparison star has slightly different parameters to the star we compare to, some scatter is introduced as a systematic offset to the determined excess fluxes of that star (see also Sec. \ref{sec:wrongstellarparas}). In fairness, such scatter will also be introduced from an incorrect set of stellar parameters when using a model. It is possible to vary the stellar parameters in use to fit the model spectra so both spectra agree in the wings, this approach can not eliminate those errors completely. The best way to handle these errors is to use as many stars as possible, as those errors will then broaden the distribution, but should not affect the resulting fit by much.

\subsection{Outliers}
\label{sec:outliers}

Fifteen objects for which observations were available had to be removed from further analysis as they could not be handled adequately with our approach.\\
- \object{HD\,114378}:
The comparison shows that the line shape differs to the one from the comparison spectrum, resulting in what appears to be a well-defined excess flux. However, this object has been found to be a binary star \citep{Malkov2012}. The observed line in the spectrum is then a combination of two (shifted) line profiles with different degrees of fill-in depending on their individual activity. The approach used here -- comparing with a spectrum of a single, main-sequence star -- is not appropriate for double stars and thus cannot be expected to yield correct results. For the same reason, we rejected results from other binary systems, such as HD\,106516 and HD\,133640.\\
- \object{HD\,6920}: This object is often listed as an F8V star (e.g., \citet{Hillen2012}), but has also been classified as subgiant, for example \citet{Abt1986,Fuhrmann1998, Gray2001,Anderson2012}. Should the latter classification be correct, it appears reasonable that the line profile differs from that of a main-sequence star to some degree. 
We have therefore excluded this object from further analysis.\\
- \object{HD\,25998}, \object{HD\,111456}, \object{HD\,115043}: These objects all have both a comparatively low value for $B-V$, as well as a high rotational velocity. The latter causes the excess flux to be smeared out across a wider spectral region than normal, which requires a high resolution and a very high S/N to disentangle the chromospheric excess flux from the photospheric contributions. Checking these spectra by eye shows
that this could not be done reliably, so we excluded results from these stars. To determine their excess flux, follow-up observations with higher S/N are needed.\\[2ex]
Removing these objects and an additional five stars with observations featuring too-low S/N in the lines of interest 
leaves us with a sample of 82 objects with a total of 2274 observations that are used to determine the conversion. Unfortunately, this leaves a rather small number of F-stars (nine objects with 265 observations in total). The other spectral types are not affected as much, with 46 G-stars (1419 observations) and 27 K-stars (590 observations). The lowest value of $B-V$ in the sample is changed to 0.43, the highest value is 1.18. See Table \ref{tab:dataoverview} for full details.

\section{Results}

\subsection{Flux of inactive stars}
\label{sec:firtinactive}

\begin{figure*}
\centering
\includegraphics[width=9cm]{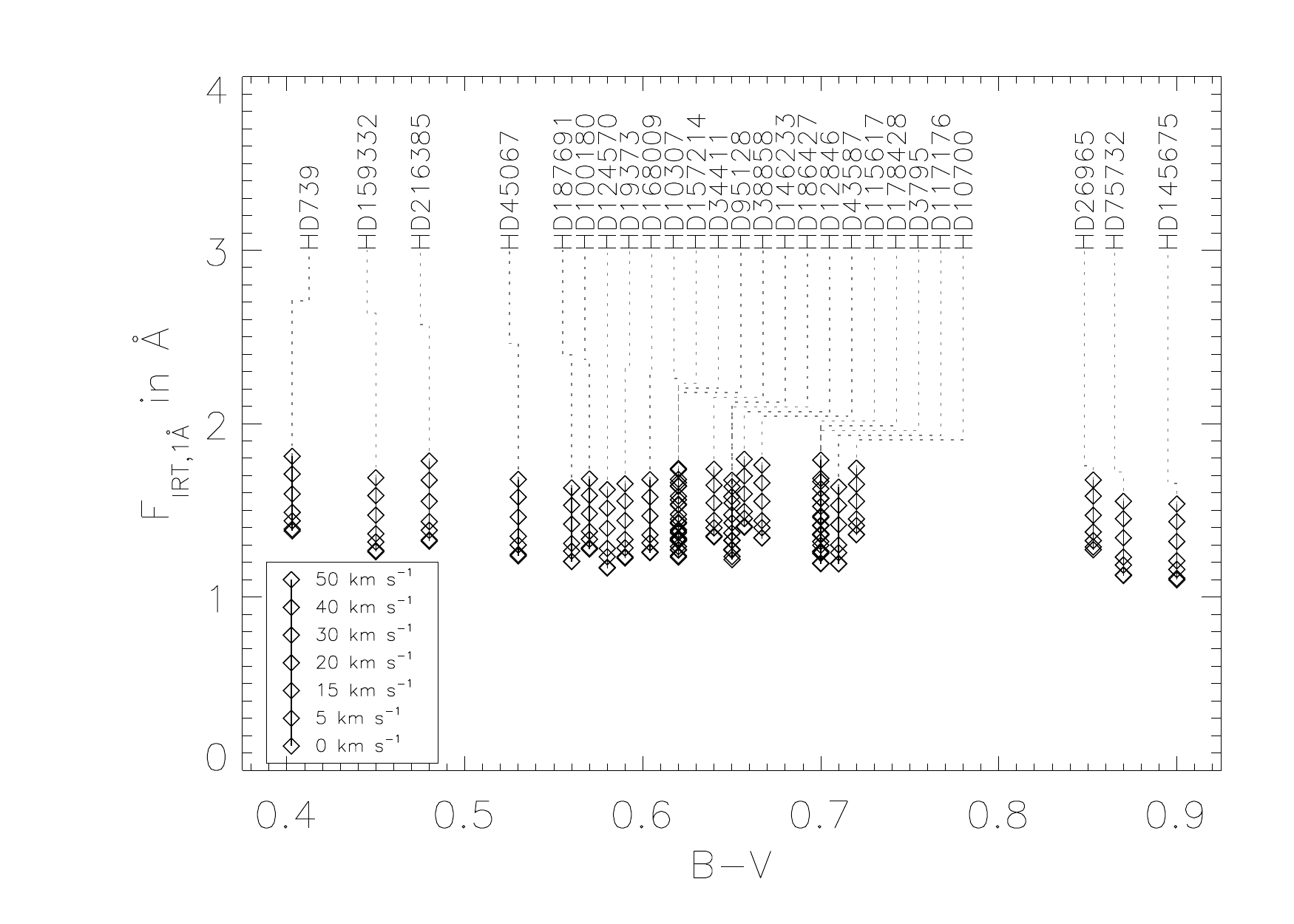}
\includegraphics[width=9cm]{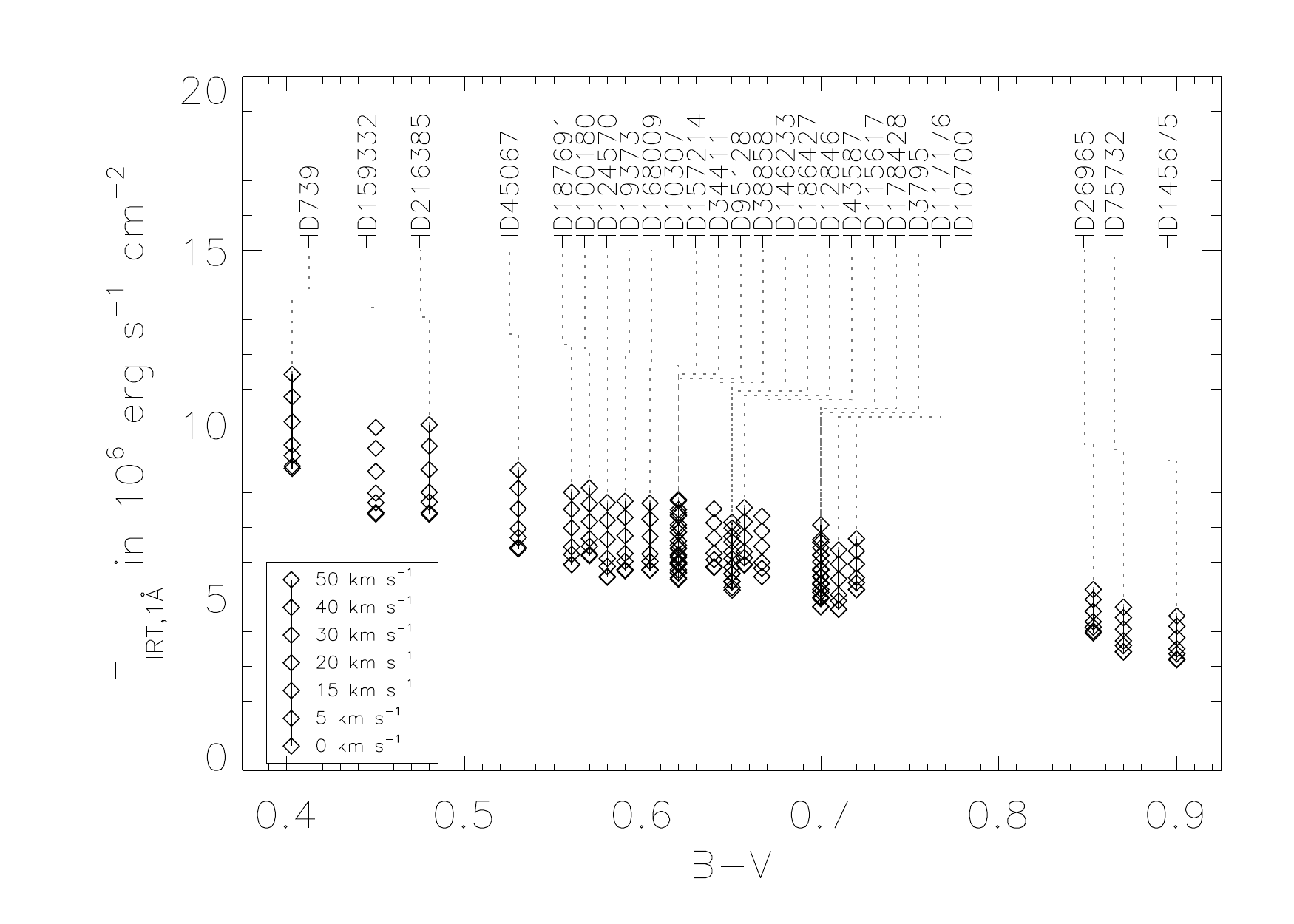}
\caption{Comparison of the obtained values for the summed-up Flux $F_{\mathrm{IRT,1\,\textup{\AA}}}$ for rotational velocities 0, 5, 15, 20, 30, 40, and 50, all in km\,s$^{-1}$ (Flux increases with higher rotational velocity). Dotted lines connect to the name of the inactive object in question. \textit{Left:} Flux in continuum units, or equivalent width in \AA. \textit{Right:} Flux converted to erg\,s$^{-1}$\,cm$^{-2}$.}
\label{fig:inactcomp}
\end{figure*}

Determining the excess flux requires a comparison spectrum to subtract the line flux of an inactive star. However, an observer may not always have a suitable spectrum at hand. In this case an estimate of the excess flux can still be performed, by calculating the inactive line flux.
Table \ref{tab:usedinactive} shows a list of inactive ($\log{R_\mathrm{HK}'}\leq4.75$), slowly-rotating stars, which we used as comparison. In this paper, we always directly subtracted their normalized spectra from the normalized spectra of the object under investigation, which also allows an independent check on the quality of the comparison by the spectra's alignment in the wings.  However, to determine the excess flux, only the (rotationally broadened) flux in the center of the line is of importance. In Fig. \ref{fig:inactcomp}, we show the resulting values for the summed-up stellar surface flux $F_{\mathrm{IRT,1\,\textup{\AA}}}$ in 1\,\AA-bandpasses for all three lines for the different inactive objects with varying simulated rotational broadening. We provide empirically derived formulae for the summed-up flux in 1\,\AA-wide bandpasses in the center of all the three \ion{Ca}{II}~IRT lines for these inactive objects as a function of rotational velocity $v\sin{i}$. The rotational broadening was performed 
according to \citet{Gray2005}. The relations are second-order 
polynomials 
fitted to the artificially rotationally broadened TIGRE spectra with a resolution of roughly 20\,000. To use these relations for determining the excess flux in a spectrum with a very different resolution, the ``bleeding'' of the flux from within the  wings due to the finite resolution must be taken into 
account.
These relations can be used to estimate the inactive line flux, and therefore to determine a value for the \ion{Ca}{ii} IRT excess flux from the spectrum of an active star. To determine this value from a spectrum, a suitable comparison star from Table \ref{tab:usedinactive} must first be found. Then, the value of $v\sin{i}$ of the observed star should be plugged into the relation given there. The result will be the summed up flux $F_{\mathrm{Inactive, IRT,1\,\textup{\AA}}}$ of all three \ion{Ca}{ii} IRT lines of this best-fitting inactive star broadened by the $v\sin{i}$ of the star under consideration, or in other words, the line flux expected if the star under consideration was inactive. This value must then be subtracted from the value obtained from an observation of an active star. We give both a relation for the converted flux $F_{\mathrm{Inactive, IRT,1\,\textup{\AA}}}$ in $10^6$~erg\,s$^{-1}$\,cm$^{-2}$, as well as for the flux in ``continuum units'', which is the integrated flux 
of a normalized spectrum in units of {\AA}. 
Subtracting the flux in continuum 
units yields an excess flux value that can be compared to $\Delta W_\mathrm{IRT}$ of \citet{Busa07}. Alternatively, this resulting value in \AA~can then be converted to physical units (erg\,s$^{-1}$\,cm$^{-2}$), for example using the relation in \citet{Hall96}, which tends to be more reliable than directly comparing values in erg\,s$^{-1}$\,cm$^{-2}$, as no error is introduced due to 
different values for $B-V$ of comparison and analyzed object.

\begin{table*}
\centering
\caption{Formulae to estimate the summed up flux in an 1\,\AA-window in the center of all three \ion{Ca}{II} IRT lines for inactive objects. $\mathrm{v}_\mathrm{rot}$ must be entered in units of km\,s$^{-1}$. To obtain the best fit, it is recommended to compare the values from normalized spectra and to subtract an additional 20\,m\AA, as described in the text. Values for $B-V$, $\log g$ and [Fe/H] are taken from \citet{Soubiran2010}}
\label{tab:usedinactive}
\begin{tabular}{lllllll}
\hline
\hline
\cline{6-7}
   &   &   &   &   & \multicolumn{2}{c}{\textbf{Estimated total flux in 1\,\AA-bandpasses in all \ion{Ca}{II}~IRT lines}}\\
\textbf{Object}  & $B-V$ & $\log g$ & [Fe/H] & $\log{R_\mathrm{HK}'}$ & in $10^{6}$ erg\,s$^{-1}$\,cm$^{-2}$ & from normalized spectra in \AA\\
 \hline \\[-1ex]
 
HD\,739 & 0.40 & 4.27 & -0.09 & -4.91 & $8.645 + 0.026\cdot v_\mathrm{rot} +6.106\cdot 10^{-4} v_\mathrm{rot}^2$&$1.371 + 0.004\cdot v_\mathrm{rot} +0.972\cdot 10^{-4} v_\mathrm{rot}^2$\\[0.8ex]
HD\,159332 & 0.45 & 3.85 & -0.23 & -4.99 & $7.326 + 0.024\cdot v_\mathrm{rot} +5.770\cdot 10^{-4} v_\mathrm{rot}^2$&$1.250 + 0.004\cdot v_\mathrm{rot} +0.987\cdot 10^{-4} v_\mathrm{rot}^2$\\[0.8ex]
HD\,216385 & 0.48 & 3.95 & -0.29 & -4.98 & $7.318 + 0.025\cdot v_\mathrm{rot} +5.791\cdot 10^{-4} v_\mathrm{rot}^2$&$1.309 + 0.005\cdot v_\mathrm{rot} +1.039\cdot 10^{-4} v_\mathrm{rot}^2$\\[0.8ex]
HD\,45067 & 0.53 & 4.01 & -0.09 & -4.90 & $6.323 + 0.025\cdot v_\mathrm{rot} +4.679\cdot 10^{-4} v_\mathrm{rot}^2$&$1.224 + 0.005\cdot v_\mathrm{rot} +0.909\cdot 10^{-4} v_\mathrm{rot}^2$\\[0.8ex]
HD\,187691 & 0.56 & 4.26 & +0.10 & -4.89 & $5.862 + 0.022\cdot v_\mathrm{rot} +4.432\cdot 10^{-4} v_\mathrm{rot}^2$&$1.190 + 0.004\cdot v_\mathrm{rot} +0.903\cdot 10^{-4} v_\mathrm{rot}^2$\\[0.8ex]
HD\,100180 & 0.57 & 4.25 & -0.06 & -4.76 & $6.134 + 0.020\cdot v_\mathrm{rot} +4.188\cdot 10^{-4} v_\mathrm{rot}^2$&$1.265 + 0.004\cdot v_\mathrm{rot} +0.866\cdot 10^{-4} v_\mathrm{rot}^2$\\[0.8ex]
HD\,124570 & 0.58 & 4.05 & +0.08 & -5.05 & $5.504 + 0.023\cdot v_\mathrm{rot} +4.536\cdot 10^{-4} v_\mathrm{rot}^2$&$1.153 + 0.005\cdot v_\mathrm{rot} +0.953\cdot 10^{-4} v_\mathrm{rot}^2$\\[0.8ex]
HD\,19373 & 0.59 & 4.21 & +0.08 & -4.84 & $5.698 + 0.021\cdot v_\mathrm{rot} +4.367\cdot 10^{-4} v_\mathrm{rot}^2$&$1.213 + 0.004\cdot v_\mathrm{rot} +0.931\cdot 10^{-4} v_\mathrm{rot}^2$\\[0.8ex]
HD\,168009 & 0.60 & 4.23 & -0.01 & -4.77 & $5.711 + 0.019\cdot v_\mathrm{rot} +4.289\cdot 10^{-4} v_\mathrm{rot}^2$&$1.242 + 0.004\cdot v_\mathrm{rot} +0.935\cdot 10^{-4} v_\mathrm{rot}^2$\\[0.8ex]
HD\,10307 & 0.62 & 4.32 & +0.03 & -4.84 & $5.635 + 0.019\cdot v_\mathrm{rot} +3.969\cdot 10^{-4} v_\mathrm{rot}^2$&$1.257 + 0.004\cdot v_\mathrm{rot} +0.888\cdot 10^{-4} v_\mathrm{rot}^2$\\[0.8ex]
HD\,157214 & 0.62 & 4.31 & -0.40 & -4.80 & $5.922 + 0.018\cdot v_\mathrm{rot} +3.963\cdot 10^{-4} v_\mathrm{rot}^2$&$1.321 + 0.004\cdot v_\mathrm{rot} +0.886\cdot 10^{-4} v_\mathrm{rot}^2$\\[0.8ex]
HD\,34411 & 0.62 & 4.22 & +0.08 & -4.85 & $5.458 + 0.021\cdot v_\mathrm{rot} +3.975\cdot 10^{-4} v_\mathrm{rot}^2$&$1.218 + 0.005\cdot v_\mathrm{rot} +0.890\cdot 10^{-4} v_\mathrm{rot}^2$\\[0.8ex]
HD\,95128 & 0.62 & 4.30 & +0.01 & -4.85 & $5.888 + 0.018\cdot v_\mathrm{rot} +4.288\cdot 10^{-4} v_\mathrm{rot}^2$&$1.312 + 0.004\cdot v_\mathrm{rot} +0.957\cdot 10^{-4} v_\mathrm{rot}^2$\\[0.8ex]
HD\,38858 & 0.64 & 4.48 & -0.22 & -4.79 & $5.798 + 0.017\cdot v_\mathrm{rot} +3.754\cdot 10^{-4} v_\mathrm{rot}^2$&$1.335 + 0.004\cdot v_\mathrm{rot} +0.868\cdot 10^{-4} v_\mathrm{rot}^2$\\[0.8ex]
HD\,146233 & 0.65 & 4.42 & +0.03 & -4.75 & $5.380 + 0.018\cdot v_\mathrm{rot} +3.774\cdot 10^{-4} v_\mathrm{rot}^2$&$1.258 + 0.004\cdot v_\mathrm{rot} +0.885\cdot 10^{-4} v_\mathrm{rot}^2$\\[0.8ex]
HD\,186427 & 0.65 & 4.32 & +0.07 & -4.80 & $5.166 + 0.019\cdot v_\mathrm{rot} +3.785\cdot 10^{-4} v_\mathrm{rot}^2$&$1.208 + 0.004\cdot v_\mathrm{rot} +0.888\cdot 10^{-4} v_\mathrm{rot}^2$\\[0.8ex]
HD\,12846 & 0.66 & 4.38 & -0.26 & -4.78 & $5.875 + 0.015\cdot v_\mathrm{rot} +3.984\cdot 10^{-4} v_\mathrm{rot}^2$&$1.391 + 0.004\cdot v_\mathrm{rot} +0.946\cdot 10^{-4} v_\mathrm{rot}^2$\\[0.8ex]
HD\,43587 & 0.67 & 4.29 & -0.04 & -4.80 & $5.528 + 0.018\cdot v_\mathrm{rot} +3.879\cdot 10^{-4} v_\mathrm{rot}^2$&$1.327 + 0.004\cdot v_\mathrm{rot} +0.933\cdot 10^{-4} v_\mathrm{rot}^2$\\[0.8ex]
HD\,115617 & 0.70 & 4.39 & -0.01 & -4.80 & $4.934 + 0.017\cdot v_\mathrm{rot} +3.450\cdot 10^{-4} v_\mathrm{rot}^2$&$1.248 + 0.004\cdot v_\mathrm{rot} +0.874\cdot 10^{-4} v_\mathrm{rot}^2$\\[0.8ex]
HD\,178428 & 0.70 & 4.25 & +0.14 & -4.88 & $4.659 + 0.018\cdot v_\mathrm{rot} +3.663\cdot 10^{-4} v_\mathrm{rot}^2$&$1.178 + 0.005\cdot v_\mathrm{rot} +0.929\cdot 10^{-4} v_\mathrm{rot}^2$\\[0.8ex]
HD\,3795 & 0.70 & 3.91 & -0.63 & -4.83 & $5.331 + 0.015\cdot v_\mathrm{rot} +4.096\cdot 10^{-4} v_\mathrm{rot}^2$&$1.347 + 0.004\cdot v_\mathrm{rot} +1.037\cdot 10^{-4} v_\mathrm{rot}^2$\\[0.8ex]
HD\,117176 & 0.71 & 3.97 & -0.06 & -4.90 & $4.584 + 0.018\cdot v_\mathrm{rot} +3.628\cdot 10^{-4} v_\mathrm{rot}^2$&$1.177 + 0.005\cdot v_\mathrm{rot} +0.934\cdot 10^{-4} v_\mathrm{rot}^2$\\[0.8ex]
HD\,10700 & 0.72 & 4.48 & -0.50 & -4.75 & $5.163 + 0.014\cdot v_\mathrm{rot} +3.325\cdot 10^{-4} v_\mathrm{rot}^2$&$1.347 + 0.004\cdot v_\mathrm{rot} +0.869\cdot 10^{-4} v_\mathrm{rot}^2$\\[0.8ex]
HD\,26965 & 0.85 & 4.51 & -0.27 & -4.89 & $3.944 + 0.012\cdot v_\mathrm{rot} +2.855\cdot 10^{-4} v_\mathrm{rot}^2$&$1.265 + 0.004\cdot v_\mathrm{rot} +0.919\cdot 10^{-4} v_\mathrm{rot}^2$\\[0.8ex]
HD\,75732 & 0.87 & 4.41 & +0.28 & -4.84 & $3.372 + 0.014\cdot v_\mathrm{rot} +2.740\cdot 10^{-4} v_\mathrm{rot}^2$&$1.110 + 0.005\cdot v_\mathrm{rot} +0.905\cdot 10^{-4} v_\mathrm{rot}^2$\\[0.8ex]
HD\,145675 & 0.90 & 4.45 & +0.41 & -4.80 & $3.150 + 0.013\cdot v_\mathrm{rot} +2.729\cdot 10^{-4} v_\mathrm{rot}^2$&$1.087 + 0.005\cdot v_\mathrm{rot} +0.944\cdot 10^{-4} v_\mathrm{rot}^2$
 \\
\hline
\end{tabular} 
\end{table*}

\begin{figure}
\resizebox{\hsize}{!}{\includegraphics{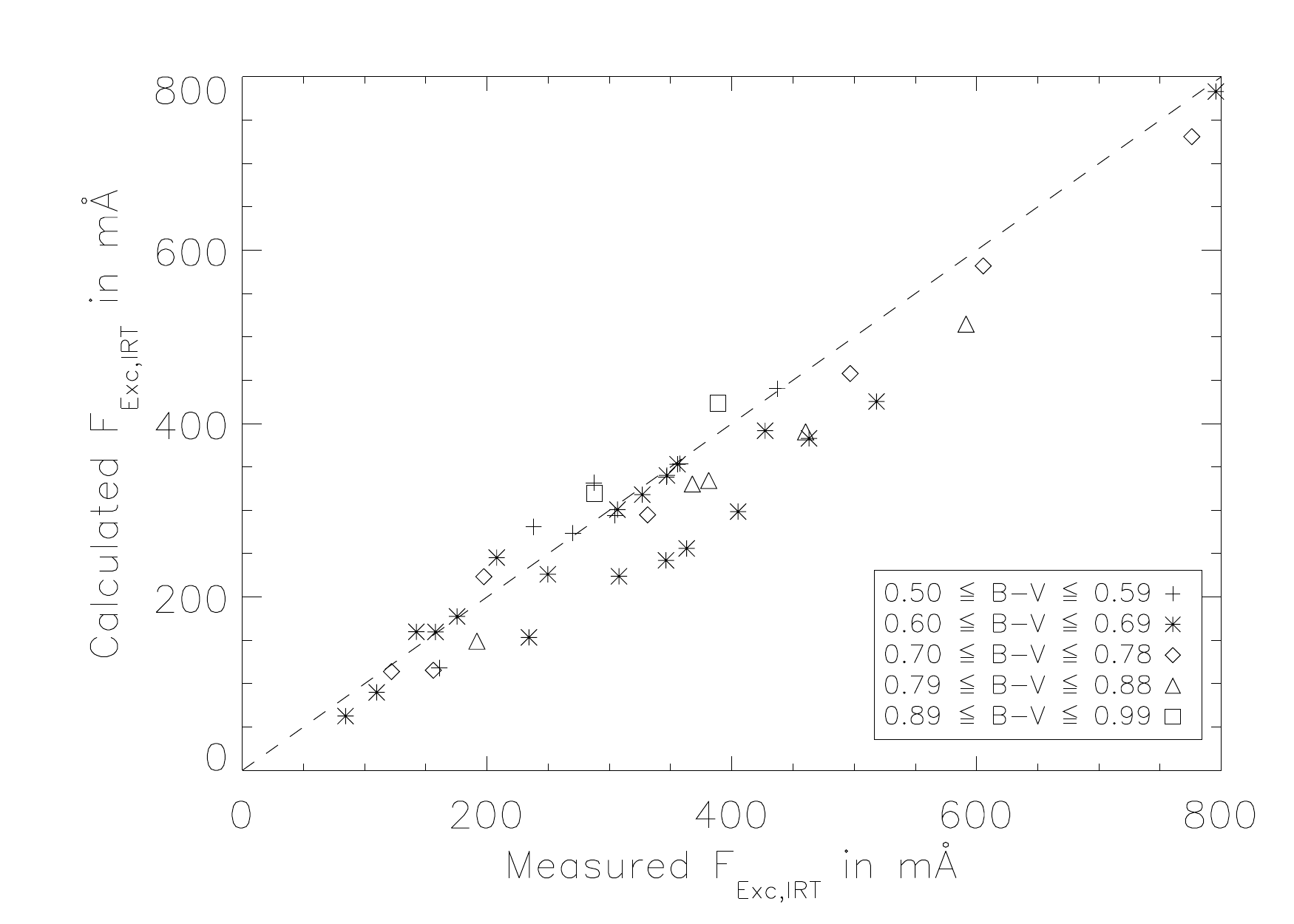}}
\caption{Comparison of the excess flux in continuum units determined using the method described in Sect. \ref{sec:comparing} to the excess flux determined from subtracting the resulting value from a fitting relation from Table \ref{tab:usedinactive} to the measured line flux, for the 42 objects with observations displaying a level of activity exceeding $R_\mathrm{HK}'= 3\cdot10^{-5}$. The dashed line corresponds to the identity relation.}
\label{fig:inactcomp_calc}
\end{figure}

Figure \ref{fig:inactcomp_calc} shows a comparison of the values measured for the excess flux using the method described in this paper against the value obtained from simply subtracting the value calculated from Table \ref{tab:usedinactive} from the measured line flux (in continuum units) for the 42 active stars with observations where $R_\mathrm{HK}' \geq 3\cdot10^{-5}$. On average, the discrepancy is about 20~m\AA, determined as the median of the residuals (68\,\% of points differ less than 45~m\AA). It should be noted that the actual measured value is on the order of 1.5~\AA, so this error is only about 1.3\,\%. Since the excess flux is the comparatively small difference of two larger values, the relative error is dramatically increased. The calculated values tend to be lower than the measured ones. The likely reason for this is the additional correction of shifting the observed spectrum to reach an agreement in the line wings to the comparison spectrum, and could thus be interpreted as the result of an 
additional correction for photospheric effects. Adding the aforementioned 20~m\AA~to the subtracted value would be a possibility to fix this.

\subsection{Measured flux in the \ion{Ca}{II} H \& K lines}
\label{sec:fhkflux}

\begin{figure}
\resizebox{\hsize}{!}{\includegraphics{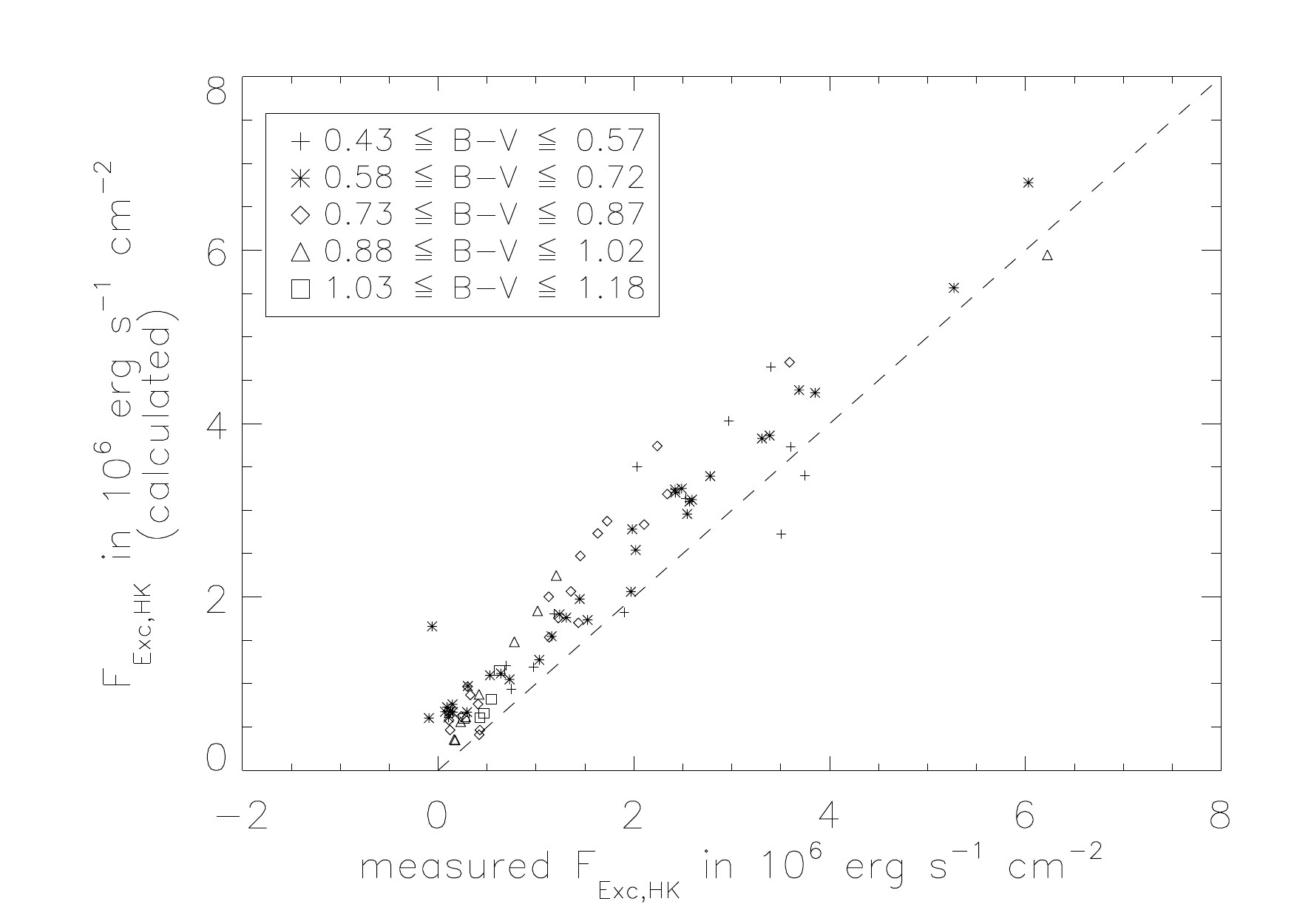}}
\resizebox{\hsize}{!}{\includegraphics{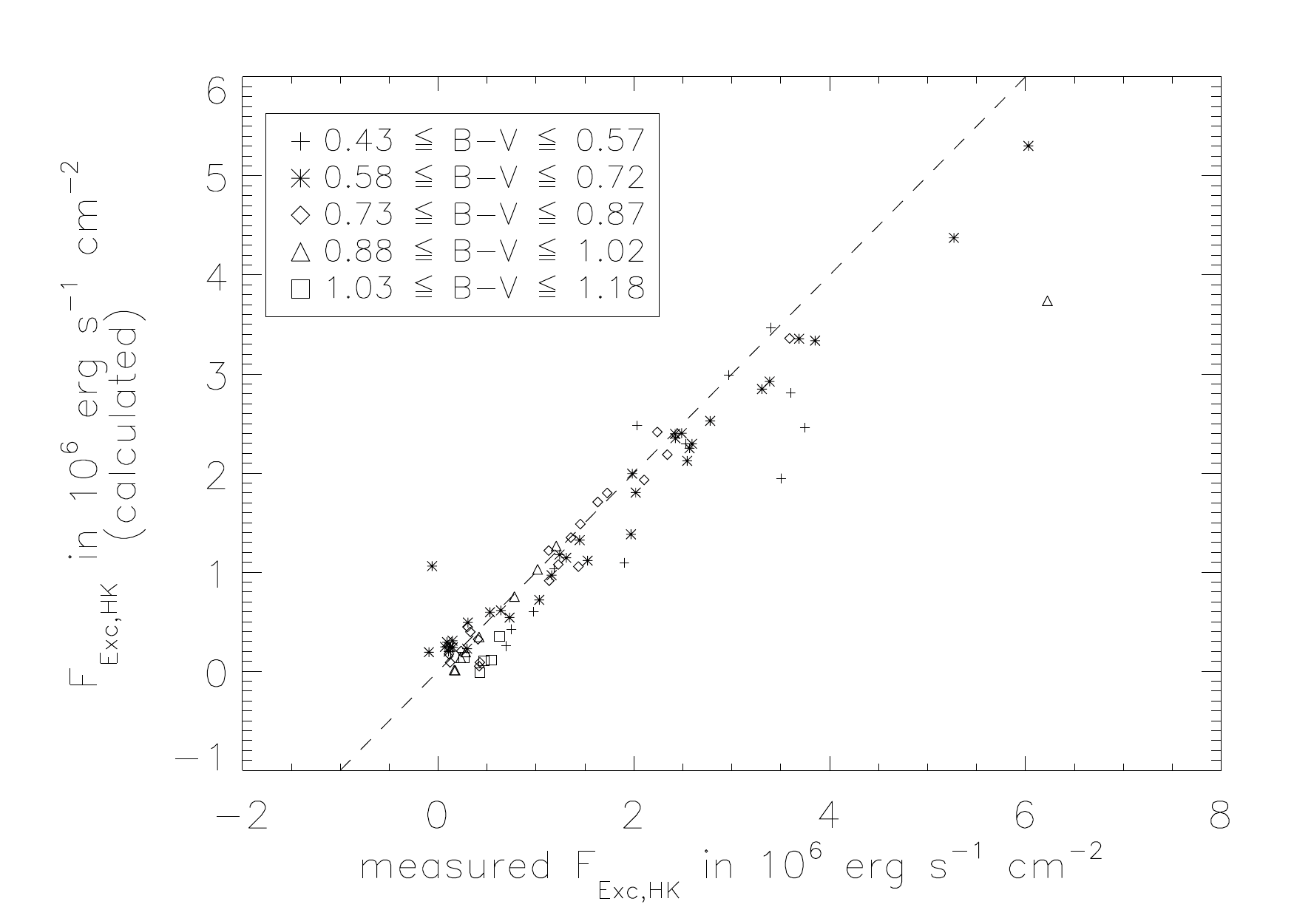}}
\caption{Comparison of the measured chromospheric excess flux in the \ion{Ca}{ii}~H\,\&\,K-lines to the one calculated from various sources. The dashed line corresponds to the identity relation. \textit{Top:} Comparing to the chromospheric excess flux according to \citet{Mittag2013} for 82 objects. \textit{Bottom:} Comparing for these same stars to the flux given in \citet{Rutten84} but corrected for photospheric and basal flux contribution, also according to \citet{Mittag2013}.}
\label{fig:fhk_comparison}
\end{figure}

\begin{figure}
\resizebox{\hsize}{!}{\includegraphics{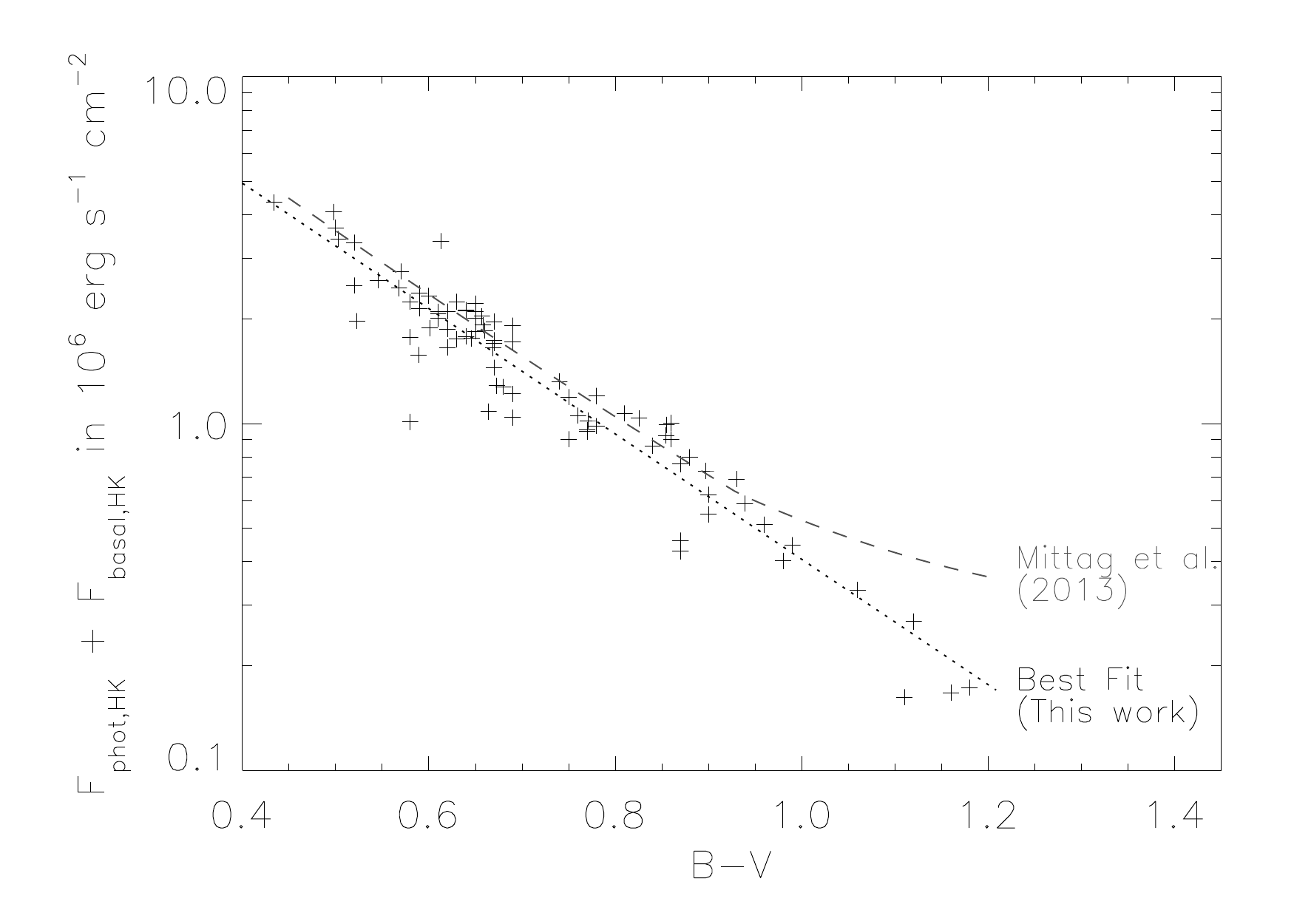}}
\caption{Comparison of the determined photospheric and basal flux from this work to the relation given in \citet{Mittag2013}. Shown are the average measured values for the 82 stars left after removal of the outliers described in Sect. \ref{sec:outliers}.} 
\label{fig:determine_basal}
\end{figure}

The procedure mentioned was also carried out for the \ion{Ca}{II} H \& K lines. While they exhibit strong changes in excess flux amplitude, the shape of these lines is very broad and contaminated by other lines, making the determination of the excess flux more difficult. Nevertheless, we obtain a measured excess flux in the center of the \ion{Ca}{II} H \& K lines, by integrating over a 2\,\AA-wide region. 
This value can be compared against the flux in the \ion{Ca}{II} H \& K lines, which can be calculated from $S_\mathrm{MWO}$ using one of several available relations in the literature. In the top plot in Fig. \ref{fig:fhk_comparison}, we compare our measured excess flux with the one calculated from the relation in \citet{Mittag2013} for the 82 stars in our sample. There, the authors present a relation for the total flux in the \ion{Ca}{II} H \& K lines, but they also give 
relations for just the photospheric flux as well as chromospheric basal flux 
contribution. Subtracting these from the total flux should in theory result in just the excess flux. Our values are lower than the calculated ones. 
As a second test, we can correct the fluxes calculated according to \citet{Rutten84} using the same relations for photospheric and basal flux contributions and compare our measurements to that result. In that comparison, the measured values are higher than the calculated ones. In \citet{Mittag2013}, the authors used PHOENIX-models to convert to fluxes in {erg\,s$^{-1}$\,cm$^{-2}$}, whereas in \citet{Rutten84}, the calibration is done using the measured solar flux. It is likely that the discrepancy originates in the different approaches of calibrating the values to physical units. To test this, we compared the total flux in the lines calculated according to \citet{Mittag2013} with the relation in \citet{Rutten84}. We find that the latter relation yields lower values, consistent with Fig. \ref{fig:fhk_comparison}.
Our measurements and the relations from the literature for the total stellar flux in the \ion{Ca}{ii}~H\,\&\,K-lines can be used to find a relation for the photospheric and basal flux for a star of given $B-V$. The relations describe a value for the total flux in the \ion{Ca}{ii}~H\,\&\,K-lines, including the photospheric and basal component, as well as the flux from activity, which we measure as excess flux: $F_\mathrm{HK} = F_\mathrm{phot, HK} + F_\mathrm{basal, HK} + F_\mathrm{Exc, HK}$, as a function of $S_\mathrm{MWO}$.
Since our measured value is $F_\mathrm{Exc, HK}$, subtracting the measured from the calculated value leaves us with just the photospheric and basal flux contributions to the line flux.
We performed this determination using the relation from \citet{Rutten84} for the total line flux (shown in Fig. 8). We could then perform a linear fit to the resulting values, and compare this relation to the one in \citet{Mittag2013}. While our relation yields lower values, the difference is not significant compared to the scatter in the datapoints for most of the covered range in $B-V$. Only for values $B-V>1.0$ do the two relations differ from one another. The relation given in \citet{Mittag2013} is defined in a step-wise fashion, and the relation changes for $B-V>0.94$. From Fig. \ref{fig:determine_basal}, it appears as if a linear extrapolation of the previous relation would result in a better fit. However, we note that our sample does not reach much further beyond this threshold value in $B-V$. 
The relation found is:
\begin{equation}
 \log{\left(F_\mathrm{phot,HK}+F_\mathrm{basal,HK}\right)} = 7.42 - 1.81 \cdot (B-V) .
\end{equation}

\subsection{Comparing measured excess fluxes of different lines}
\label{sec:diffflux}

\begin{figure*}
\centering
\includegraphics[width=9cm]{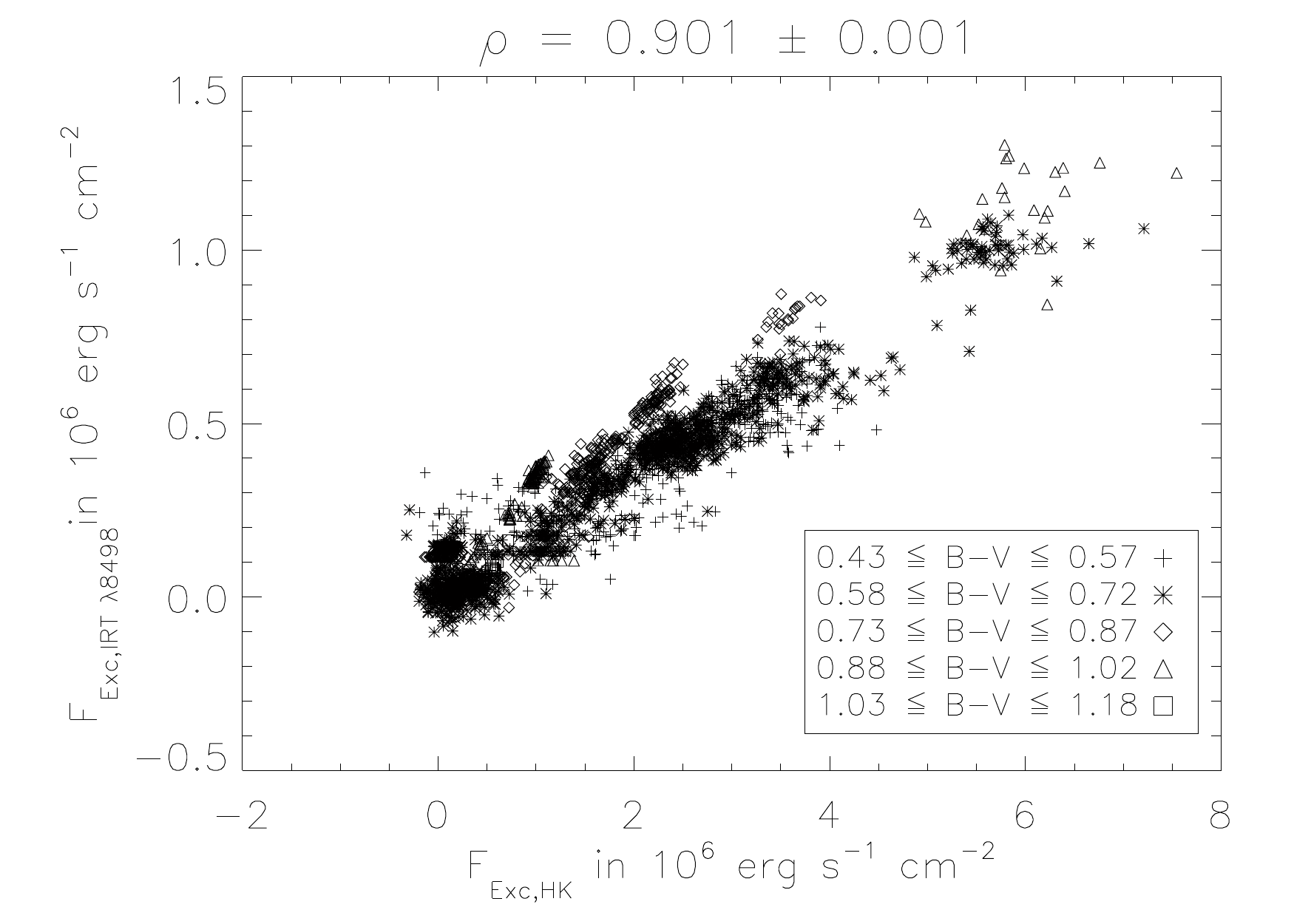}
\includegraphics[width=9cm]{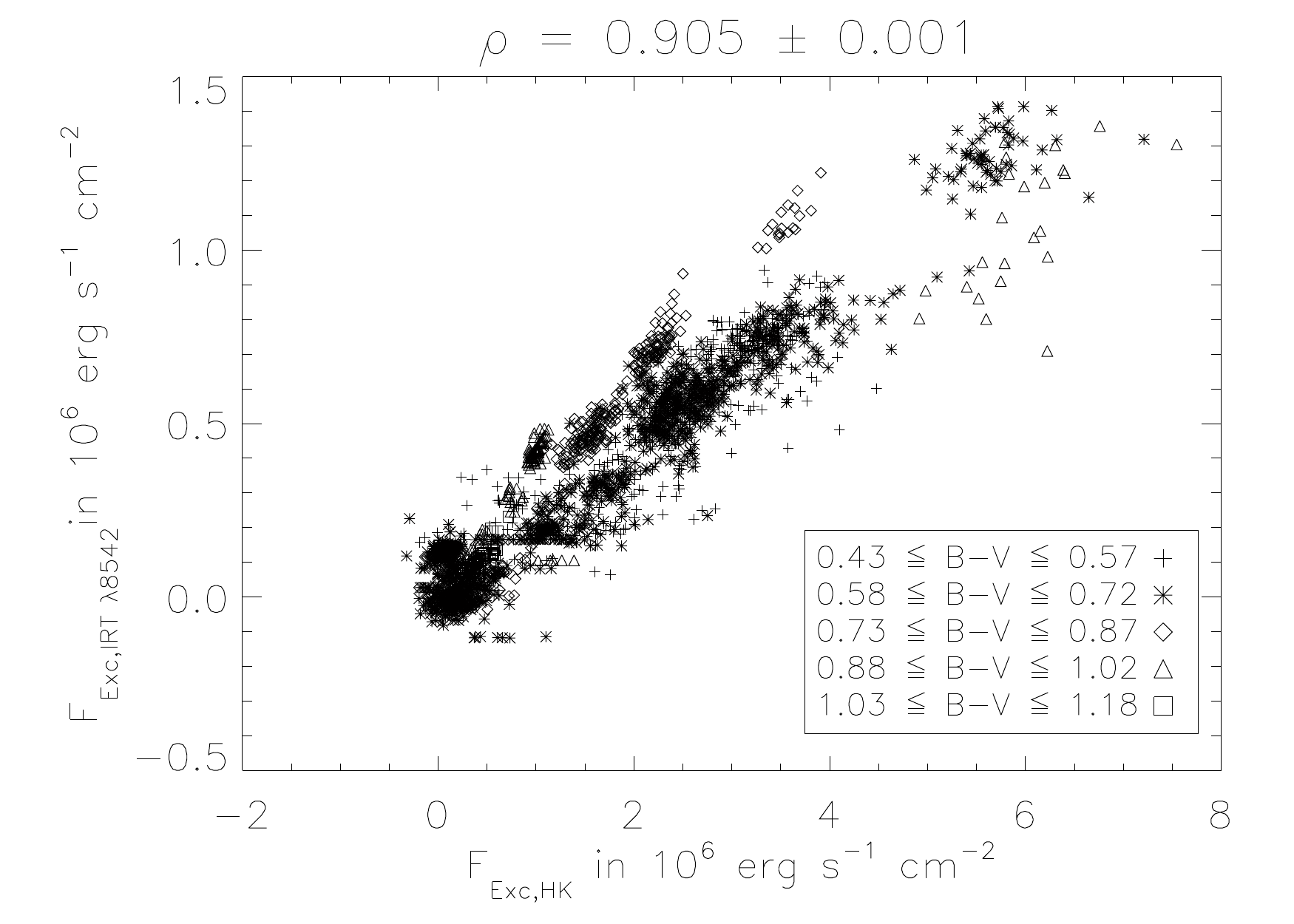}
\includegraphics[width=9cm]{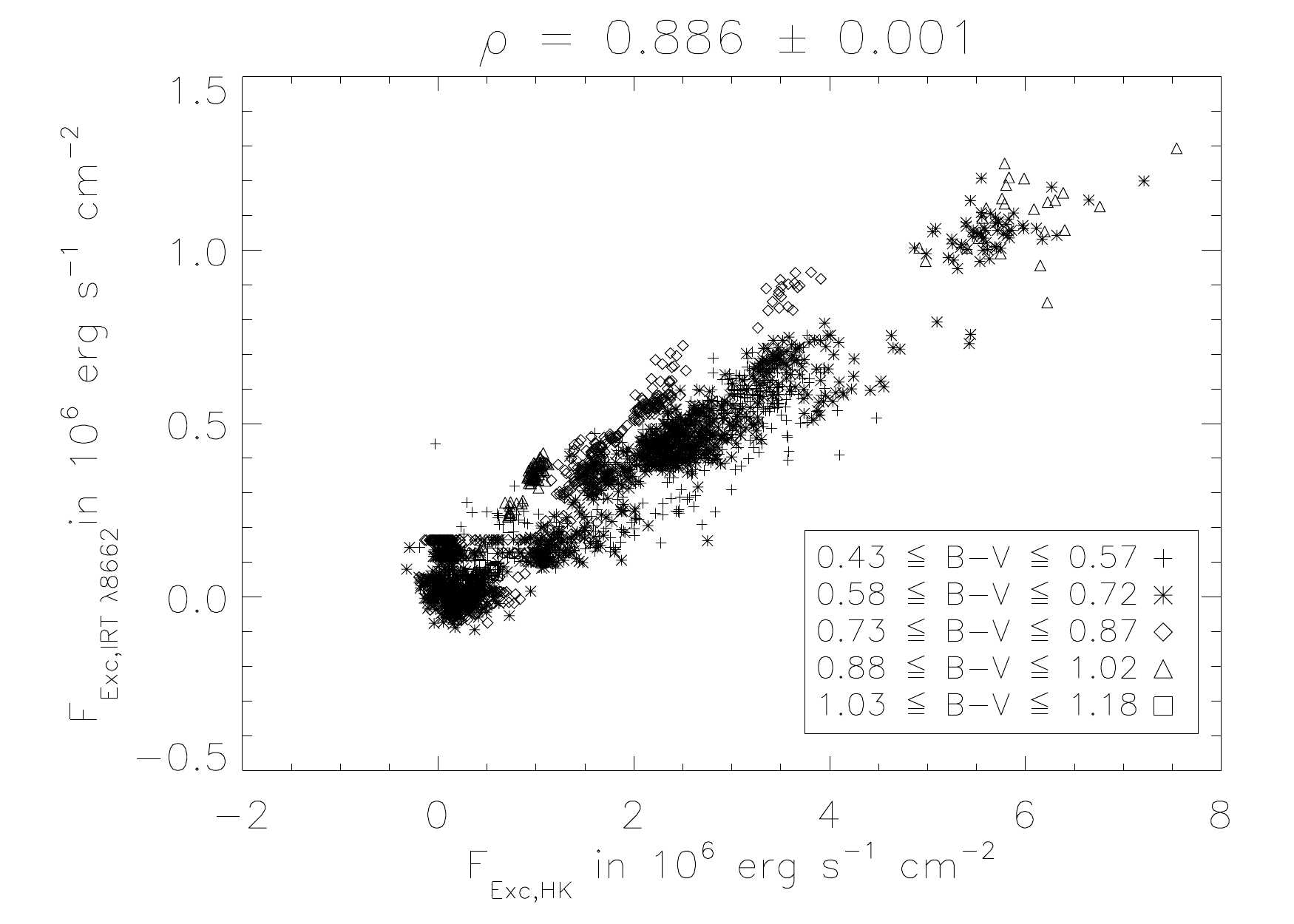}
\includegraphics[width=9cm]{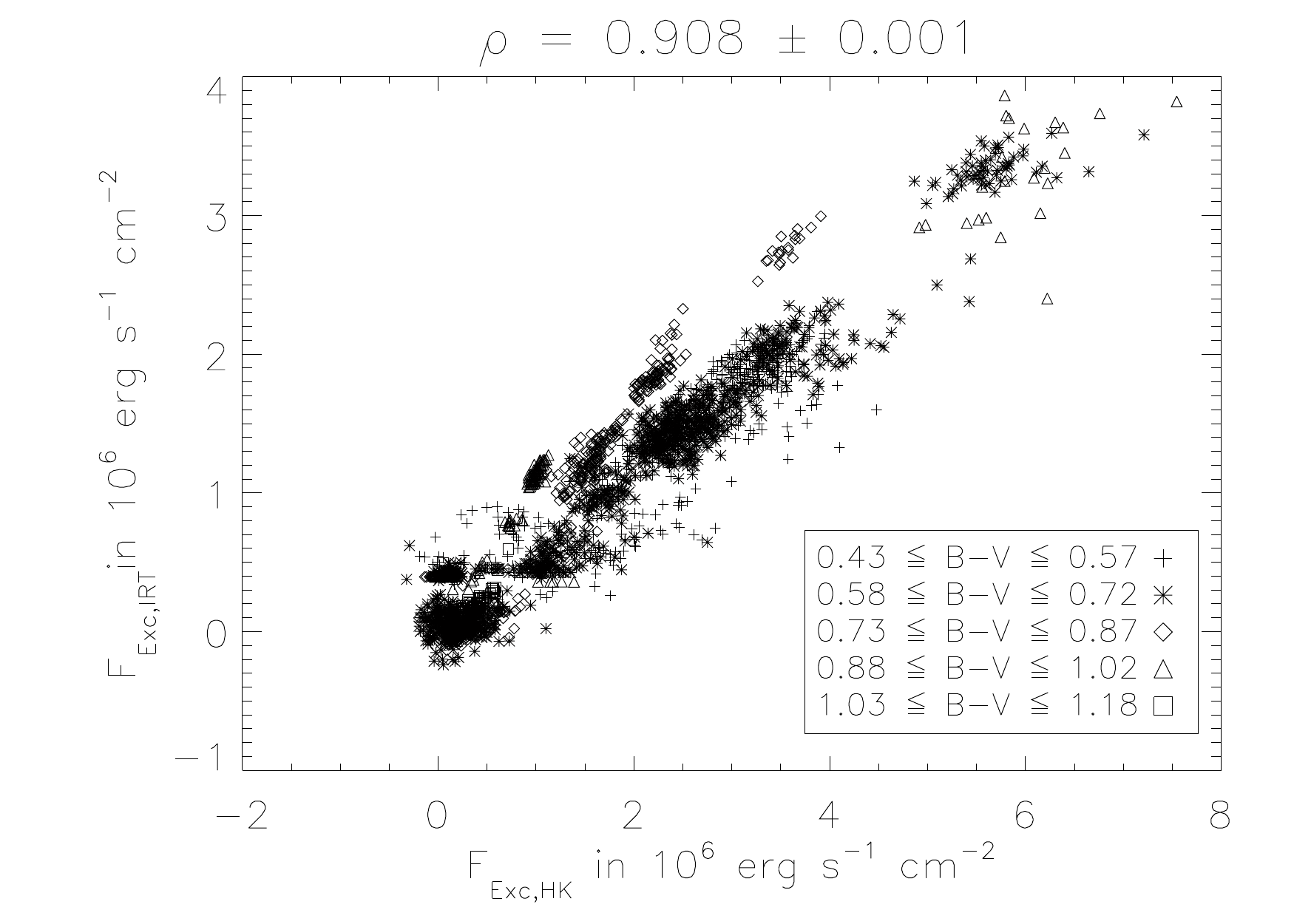}
\caption{Measured chromospheric excess flux in the \ion{Ca}{II}~H~\&~K lines compared with the excess flux in the individual \ion{Ca}{II}~IRT lines, as well as the sum of all \ion{Ca}{II}~IRT lines (bottom right). These plots include data from 2274 observations of 82 stars.}
\label{fig:fhk_vs_firt}
\end{figure*}

\begin{figure}
\resizebox{\hsize}{!}{\includegraphics{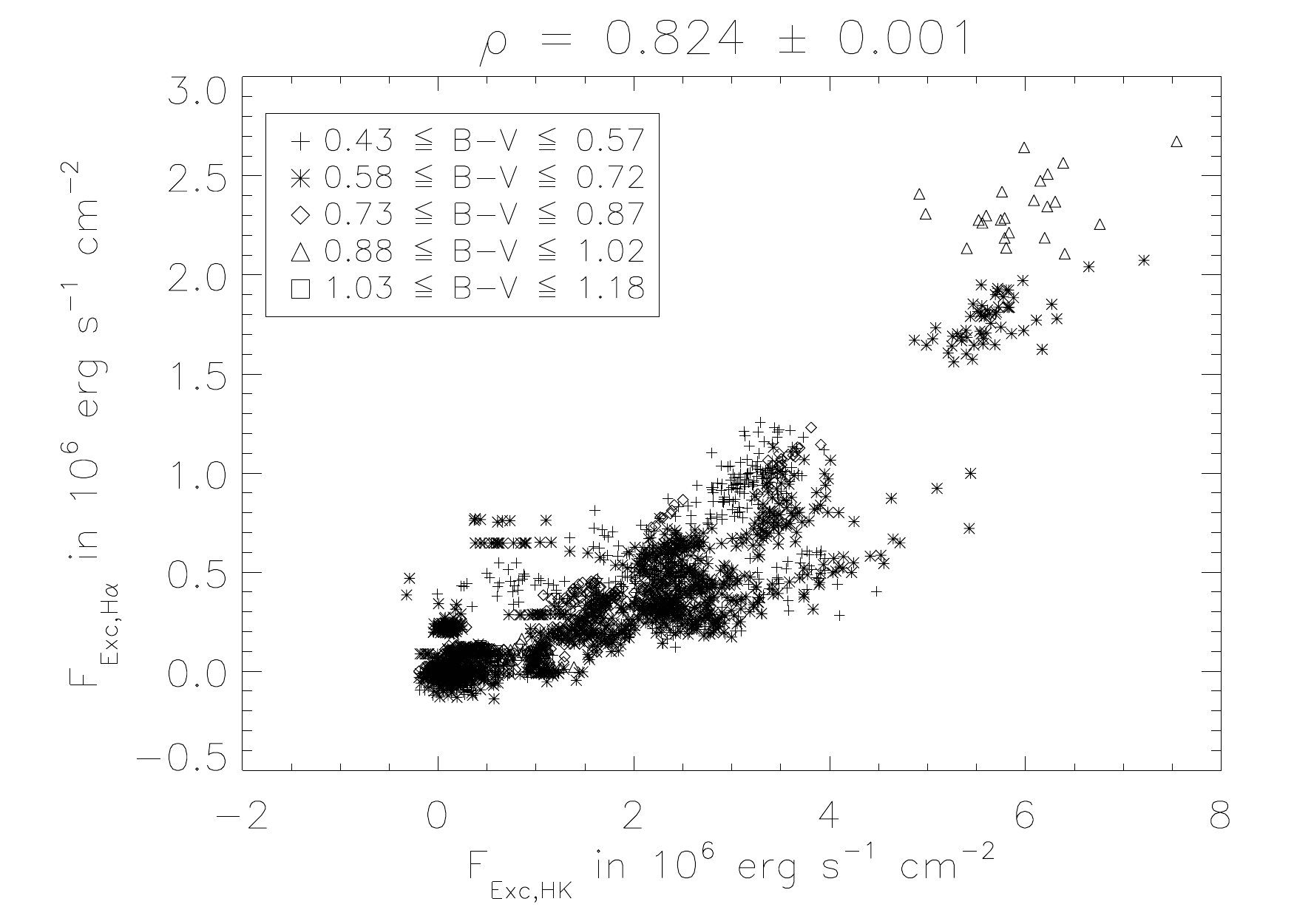}}
\caption{Measured chromospheric excess flux from 2274 observations of 82 stars in the \ion{Ca}{II} H \& K lines compared with the excess flux in H$\alpha$. Fluxes given in erg\,s$^{-1}$\,cm$^{-2}$.}
\label{fig:fhk_vs_fhalpha}
\end{figure}

We obtained 2274 values from 82 stars for the measured chromospheric flux in the \ion{Ca}{II} H, K  and IRT lines, as well as H$\alpha$, converted to real physical units by interpolating the relations in \citet{Hall96}. The determined excess fluxes do not include any photospheric or chromospheric basal flux contributions, as those have been removed by subtraction of the comparison spectrum. The resulting plots for the three individual lines in the \ion{Ca}{II}~IRT, as well as the sum of all three lines, compared to the measured flux in the \ion{Ca}{II} H \& K lines are shown in Fig. \ref{fig:fhk_vs_firt}. The second \ion{Ca}{II} IRT line shows both a strong correlation, as well as the largest fill-in, implying that it is the most sensitive line of the three. We obtain a very obvious correlation. We determined the Spearman's correlation value $\rho$ to be largest ($\rho \approx 0.908$) for the correlation between the summed-up excess flux in all three \ion{Ca}{II}~IRT lines and the excess flux in the \ion{Ca}{
II}
 H \& K 
lines.\\
H$\alpha$, another often-used indicator, also shows a correlation (Fig. \ref{fig:fhk_vs_fhalpha}), but the scatter is larger, and therefore we obtain a lower value with $\rho \approx 0.824$. It has been shown previously that H$\alpha$ does not always correlate with the \ion{Ca}{II} H \& K line indicators \citep{Cincunegui07,Meunier09,GomesdaSilva14}. Many stars in our sample show less variation in the excess fluxes than the errors on the individual measurements, so that they cannot be used to reliably estimate the correlation for an individual star. Using only the 68 stars with five or more observations for which the errors on the excess fluxes are significantly lower than their variation, we found the Spearman correlation to cover the entire range from -1.0 for some stars to 1.0 for others. The median correlation between the two excess fluxes is only $\rho\approx0.24$. In contrast, performing the same analysis for \ion{Ca}{ii} excess fluxes, the median 
correlation is $\rho\approx0.54$, 
significantly higher. Additionally, the number of stars with a negative correlation between the two excess fluxes is much lower. \\
The obtained excess fluxes in the \ion{Ca}{II} H \& K line correlate very well with each other ($\rho \approx 0.95$), and the measured flux in the K line is about 33\,\% higher than in the H line. This is similar to the value of 27\,\% observed by \citet{Wilson68}.

\subsection{Fits to the excess}
\label{sec:fittedexcess}

For each observation, we fit Gaussians to the excess flux. We then checked if parameters obtained in this way showed any correlation to known activity indices. The amplitudes of the fitted Gaussians do show a correlation to the integrated flux in the \ion{Ca}{ii}~H~\&~K lines with $\rho\approx0.7$ after removal of obvious outliers, yet the determined amplitudes have very large errors and are thus less suited for conversion than the integrated flux. Additionally, this method suffers more strongly from high noise in the spectra, as single spikes from noise can dominate the fit. This is the reason for the larger number of outliers.
The width of the fitted Gaussian shows no correlation to the integrated flux in the \ion{Ca}{ii}~H~\&~K lines, or any of the established activity indicators.

\section{Conversion relations}
\label{sec:conversion}

Because the excess fluxes in the \ion{Ca}{ii} lines and the indices derived from them are well-correlated, we can make use of our comparatively large sample size and find relations to convert one parameter into another. 
We assume that the two indices we wish to convert into one another follow a linear relation. We do however, allow the coefficients in the conversion to depend on stellar parameters. Here, we use $B-V$, but equivalently $T_\mathrm{eff}$ could also be used.
Letting $x$ be the index to be converted into another index $y$, we then set out to find the relation:
\begin{equation}
 y = m(B-V) \cdot x + b(B-V).
 \label{eq:simplelinear}
\end{equation}

If we assume $m$ and $b$ to be a polynomial, we can perform a regression to determine the coefficients. However, our data is not equally sampled in $B-V$. Therefore, if we perform the regression without taking this fact into consideration, we might find the resulting polynomial to just be optimized for the regions in $B-V$ where many stars of our sample lie in. To avoid this, we selected subsets of all datapoints. For fifteen different values of $B-V$, we select only observations of stars close to that value, and then fit Eq. \ref{eq:simplelinear} only for the datapoints from that subset of stars, which yields the values $m$, $b$ only for that specific $B-V$. Since our objects are not evenly distributed in $B-V$, our sampling in $B-V$ is not equidistant. Instead, we selected the different values for $B-V$ for which we perform this fit so there are datapoints of at least three stars for each subset. 
On average, a subset includes $\sim$190 observations and seven stars. 
Finally, we fit a polynomial to the found values $m$ and $b$, or their logarithm, for each $B-V$ sampled to obtain the relations $m(B-V)$ and $b(B-V)$. For completeness' sake, we have determined the coefficients with regression as well. The values converted by that approach are of similar quality. In the following, we will discuss the relations found in more detail for some specific pairs of observed indicators.

\subsection{Excess flux in the \ion{Ca}{II} H \& K lines computed from $S_\mathrm{MWO}$}

\begin{figure}
\resizebox{\hsize}{!}{\includegraphics{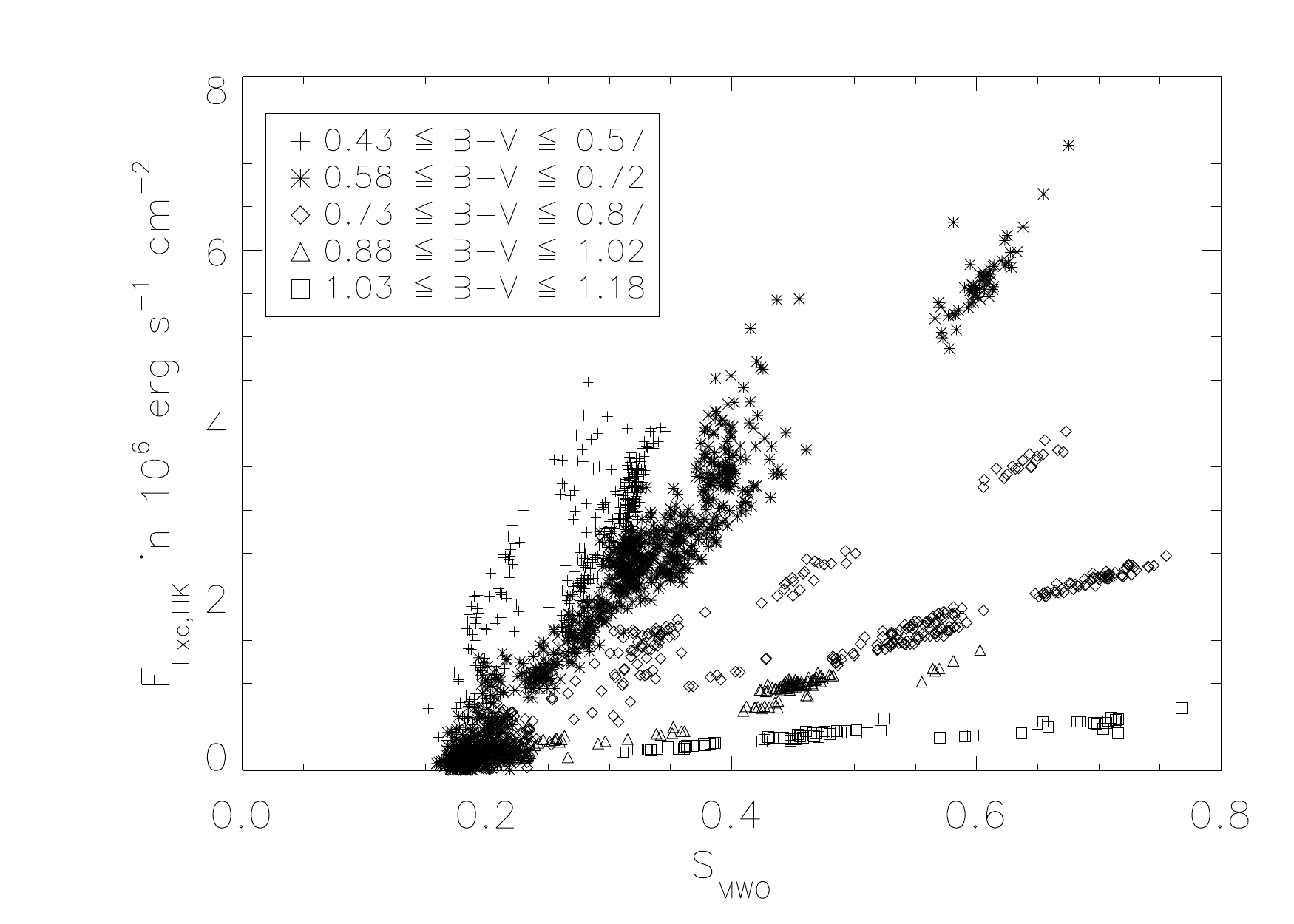}}
\caption{Measured excess flux in the \ion{Ca}{II} H \& K lines as a function of $S_\mathrm{MWO}$. This plot shows data from 2154 observations of 80 stars. For reasons of clarity, we have removed \object{HD\,22468} from this plot, as it contributes a number of datapoints clustering around a value of $S_\mathrm{MWO}\approx1.25$, as well as some observations with a negative value of $F_\mathrm{Exc,HK}$ that is consistent with zero considering its error.}
\label{fig:fhk_vs_s}
\end{figure}

Figure \ref{fig:fhk_vs_s} shows our measured excess flux compared with the corresponding value for $S_\mathrm{MWO}$, with different symbols again corresponding to different values of $B-V$. It appears that the relation between the two parameters is linear, yet the exact values of the linear fit coefficients depend on $B-V$. As previously described, we obtained relations for slope ($m$) and intercept ($b$) for individual values of $B-V$, after removal of 120 observations that were either clearly inaccurate (e.g., negative excess flux values), or too noisy with a strong influence on any linear fit performed (\object{HD\,22468}), to not have the fit dominated by noisy data. We found a second-order polynomial fit for $\log{m}$ and $b$ respectively to give good results:
\begin{eqnarray}
\label{eq:fhk_from_s}
F_\mathrm{Exc, HK} & = & (m \cdot S_\mathrm{MWO} + b)\cdot10^6\mathrm{\,erg\,s}^{-1}\mathrm{cm}^{-2},\mathrm{~with}\\
\log{m} &=& 1.027 + 1.718 \cdot (B-V) - 2.440 \cdot (B-V)^{2} \nonumber \\
b &=& -2.908 - 0.667 \cdot (B-V) + 3.249 \cdot (B-V)^{2} \nonumber
 \end{eqnarray}

\begin{figure}
\resizebox{\hsize}{!}{\includegraphics{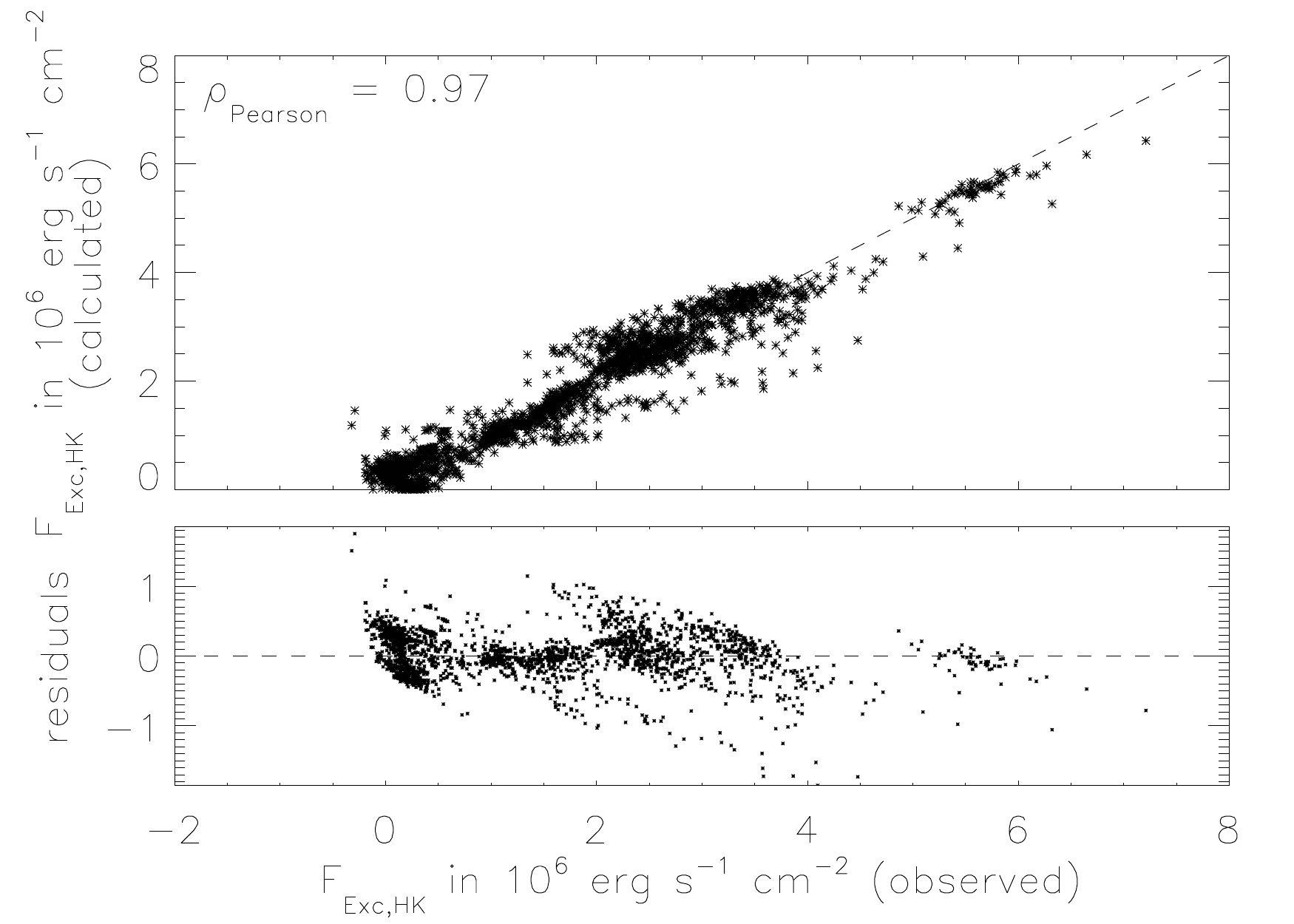}}
\caption{Comparison of the measured excess flux in the \ion{Ca}{II} H \& K lines to the one converted from $S_\mathrm{MWO}$ using Eq. \ref{eq:fhk_from_s}. The dashed line is the identity (\textit{Top}), or the zero-level (\textit{Bottom}). This plot shows data from 2137 observations of 79 stars. After conversion, the Pearson correlation coefficient is 0.97, indicating a strong linear correlation.}
\label{fig:fhk_from_s}
\end{figure}

As this formula was found using data from only F, G and K main sequence stars, it is only valid for those, with a valid $B-V$ ranging from $\sim0.5$ to $\sim1.0$. 
Figure \ref{fig:fhk_from_s} compares the converted value from $S_\mathrm{MWO}$ to the measured value. We can estimate the error of the converted values from the average of the residuals to be $3.0\cdot10^5$\,erg\,s$^{-1}$\,cm$^{-2}$. This value is not a true 1\,$\sigma$-value, however, as the distribution is not Gaussian (68\,\% lie within $3.0\cdot10^5$\,erg\,s$^{-1}$\,cm$^{-2}$, 95\,\% within $7.0\cdot10^5$\,erg\,s$^{-1}$\,cm$^{-2}$). The stated error corresponds to an average relative error of about 11\,\%. The quality of the conversion can be estimated from the Pearson correlation coefficient, which is a measure on the linear correlation. Here, we find $\rho_\mathrm{Pearson} = 0.97$, indicating that the conversion worked well, as expected.\\

\subsection{Flux conversion}
\label{sec:fluxconv}

\begin{figure}
\resizebox{\hsize}{!}{\includegraphics{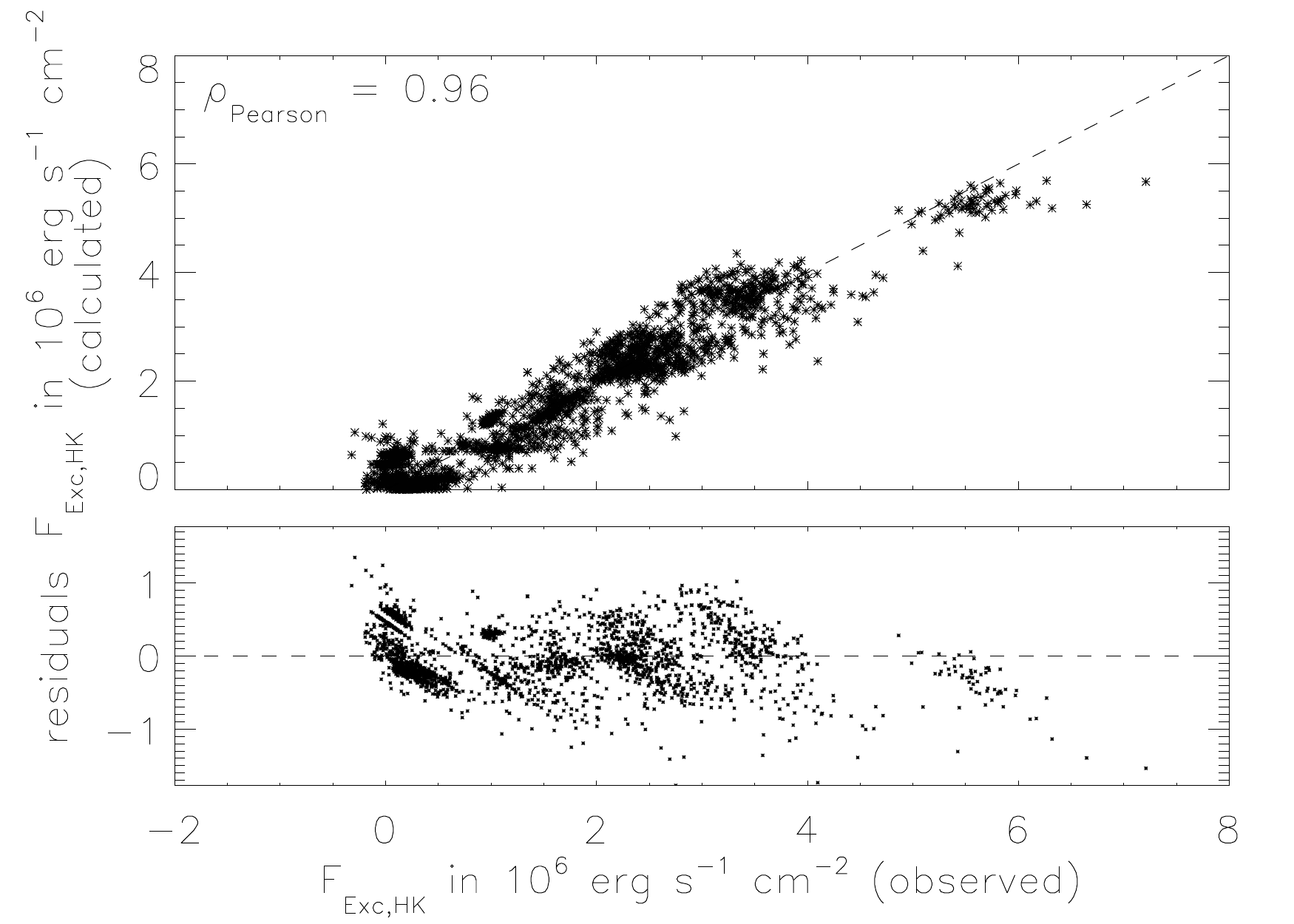}}
\caption{Comparison of the measured values of the excess flux in the \ion{Ca}{ii} H \& K lines with the one converted from the excess flux in the \ion{Ca}{ii}~IRT lines using Eq. \ref{eq:irt-to-hk}. This plot includes data from 2234 observations of 80 stars. The dashed line corresponds to the identity relation \textit{(top)}, or the zero-level \textit{(bottom)}.}
\label{fig:irt_to_fhk}
\end{figure}

As shown in Sect. \ref{sec:diffflux}, the excess flux in the \ion{Ca}{II}~IRT lines is strongly correlated with the one seen in the \ion{Ca}{II} H \& K lines (Fig. \ref{fig:fhk_vs_firt}). From definition of those values, it is apparent that no intercept $b$ is needed here. We find a dependence of the logarithm of the slope $m$ to $B-V$ in our dataset. First, due to the different temperatures, the surface flux ratio at the different lines introduces a rather strong dependence, towards higher excess fluxes in the \ion{Ca}{II} H \& K lines for lower values of $B-V$. However, when using the excess flux in continuum units, this effect disappears. The resulting $B-V$-dependence now results in the opposite direction, and there is a trend towards higher excess fluxes for higher $B-V$, evident from the different sign in the slope relation below (Eq. \ref{eq:irt-to-hk} and Eq. \ref{eq:irt-to-hk-cont}). Of those two effects, the surface flux ratio at the different points in the 
continuum is 
larger and thus 
dominates. For our sample of stars, 
relations linear in $B-V$ result in a good fit. 
To convert fluxes in erg\,s$^{-1}$\,cm$^{-2}$:
\begin{eqnarray}
\label{eq:irt-to-hks}
F_\mathrm{Exc,HK} & = & 10^{1.095 - 0.587 \cdot (B-V)} \cdot F_{\mathrm{Exc,IRT\,}\lambda8498} \\
F_\mathrm{Exc,HK} & = & 10^{1.036 - 0.631 \cdot (B-V)} \cdot F_{\mathrm{Exc,IRT\,}\lambda8542} \nonumber \\
F_\mathrm{Exc,HK} & = & 10^{1.137 - 0.663 \cdot (B-V)} \cdot F_{\mathrm{Exc,IRT\,}\lambda8662} \nonumber \\
\label{eq:irt-to-hk}
F_\mathrm{Exc,HK} & = & 10^{0.606 - 0.612 \cdot (B-V)} \cdot F_\mathrm{Exc,IRT} 
 \end{eqnarray}
And to convert normalized fluxes in continuum units:
\begin{equation}
\label{eq:irt-to-hk-cont}
F_\mathrm{Exc,HK} = \left(-0.085 + 1.402 \cdot (B-V)\right) \cdot F_\mathrm{Exc,IRT}
 \end{equation}
As before, these relations are, by nature of how they were determined, only valid for F, G and K main-sequence stars with $B-V$ ranging from $\sim0.5$ to $\sim1.0$. The Pearson correlation coefficient of the converted to observed values is larger than 0.95 in all cases. \linebreak
The errors of such a conversion have been estimated from the residuals to be about $4\cdot10^5$\,erg\,s$^{-1}$\,cm$^{-2}$ and 60\,m\AA, respectively. In Fig. \ref{fig:irt_to_fhk}, we compare the converted values from Eq. \ref{eq:irt-to-hk} to the measured values.\\
We note that the \ion{Ca}{II}~IRT lines are well correlated with each other. Therefore, measuring the excess flux in just one allows estimating it in the others from a linear relation. Equations for such a conversion are shown in Table \ref{tab:cairttoeachother}. From these parameters, it is evident that the second line is the most sensitive of them, with the largest fill-in observed.

\begin{table*}
\centering
\caption{Relations to estimate the excess flux in a \ion{Ca}{II}~IRT line from measurements of another. The errors of such a conversion are about 30\,000\,erg\,s$^{-1}$\,cm$^{-2}$.}
\label{tab:cairttoeachother}
\small
\begin{tabular}{rlll}
\hline
\hline
\multicolumn{1}{c}{\textbf{Source line}} & \multicolumn{3}{c}{\textbf{Target line}} \\
 \cline{2-4} \\[-1.5ex]
  & \textbf{\ion{Ca}{II}~IRT $\lambda8498$} & \textbf{\ion{Ca}{II}~IRT $\lambda8542$}& \textbf{\ion{Ca}{II}~IRT $\lambda8662$}\\
 \hline
 \textbf{\ion{Ca}{II}~IRT $\lambda8498$} &   & $F_\mathrm{Exc,IRT\,\lambda8542} = 1.232 \cdot F_\mathrm{Exc,IRT\,\lambda8498}$ & $F_\mathrm{Exc,IRT\,\lambda8662} = 1.006 \cdot F_\mathrm{Exc,IRT\,\lambda8498}$\\
\textbf{\ion{Ca}{II}~IRT $\lambda8542$} & $F_\mathrm{Exc,IRT\,\lambda8498} = 0.801 \cdot F_\mathrm{Exc,IRT\,\lambda8542}$ &   & $F_\mathrm{Exc,IRT\,\lambda8662} = 0.808 \cdot F_\mathrm{Exc,IRT\,\lambda8542}$\\
\textbf{\ion{Ca}{II}~IRT $\lambda8662$} & $F_\mathrm{Exc,IRT\,\lambda8498} = 0.976 \cdot F_\mathrm{Exc,IRT\,\lambda8662}$ & $F_\mathrm{Exc,IRT\,\lambda8542} = 1.210 \cdot F_\mathrm{Exc,IRT\,\lambda8662}$ & \\
\hline
\end{tabular} 
\end{table*}

\subsection{Converting \ion{Ca}{II} IRT measurements to known activity indices}

\begin{figure}
\resizebox{\hsize}{!}{\includegraphics{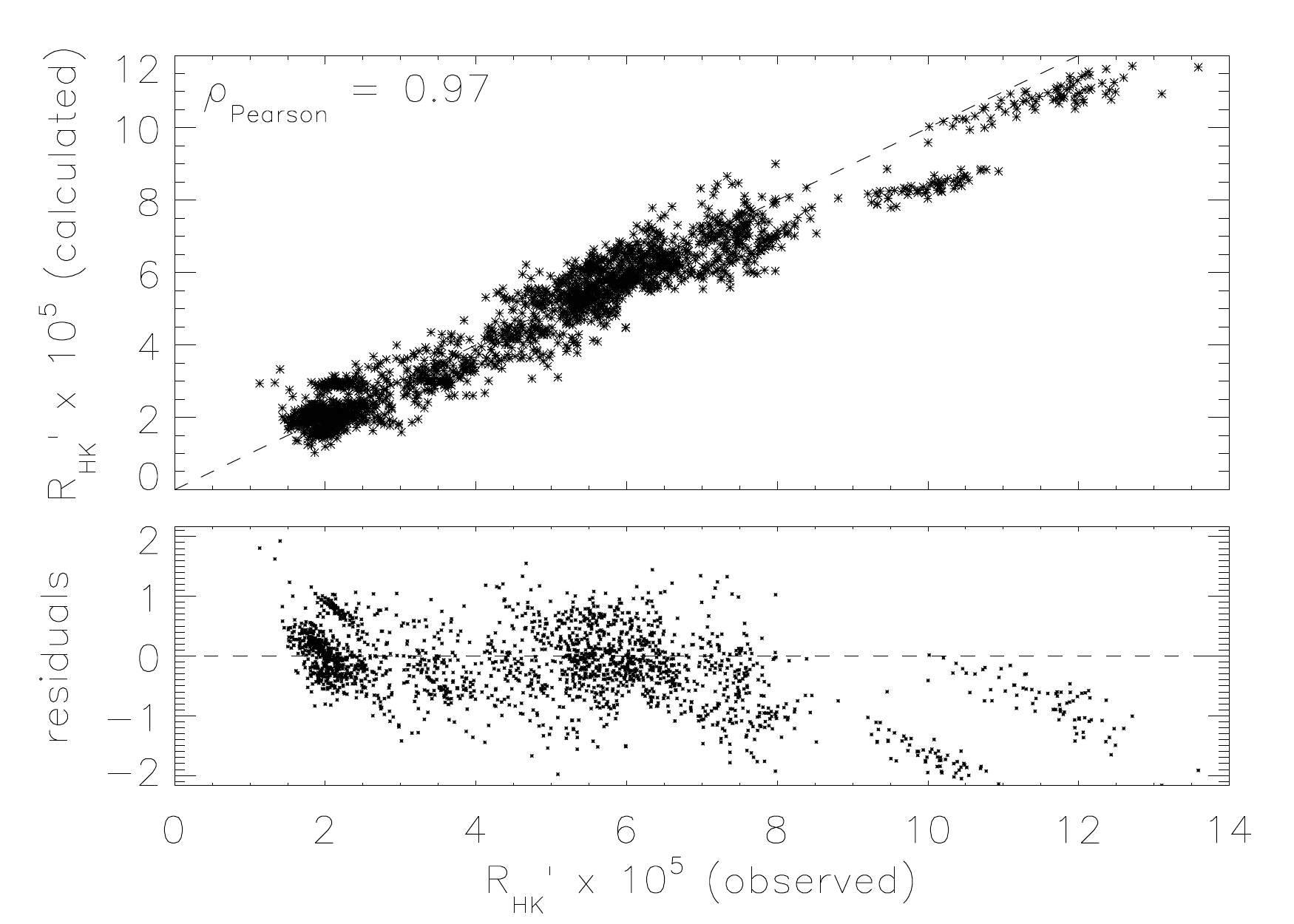}}
\resizebox{\hsize}{!}{\includegraphics{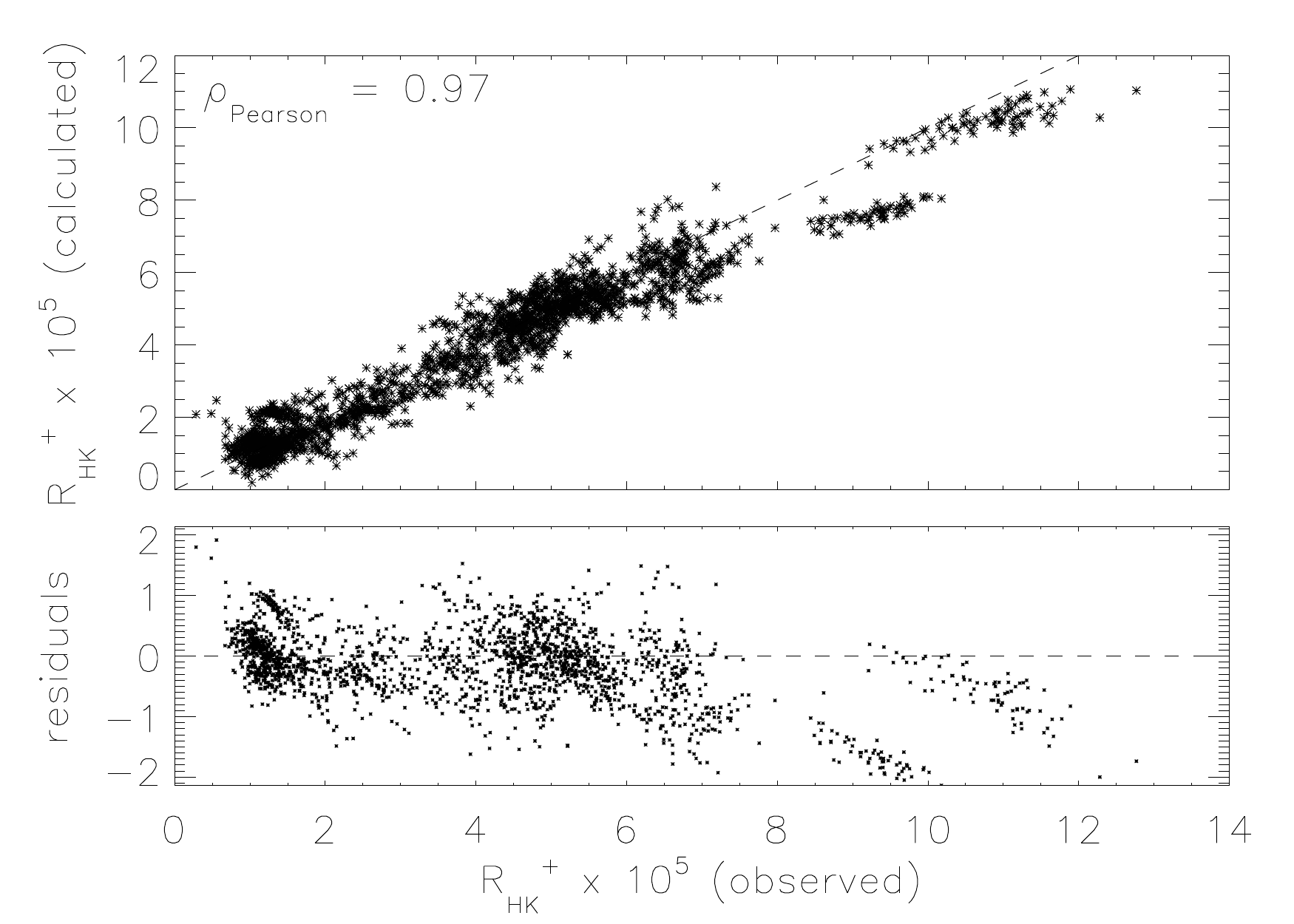}}
\caption{Comparison of the measured value for $R_\mathrm{HK}'$ (top), resp. $R_\mathrm{HK}^{+}$ (bottom) to the one converted from the excess flux in the \ion{Ca}{II}~IRT lines, using Eq. \ref{eq:rhk_from_irt} and Eq. \ref{eq:rhkplus_from_irt}. These plots include data from 2076 observations of 76 stars. The dashed line corresponds to the identity relation \textit{(top)}, or the zero-level \textit{(bottom)}. The Pearson correlation coefficient of converted to observed values is 0.97 in both cases, indicating a good conversion.}
\label{fig:RHKcompare}
\end{figure}

\begin{figure}
\resizebox{\hsize}{!}{\includegraphics{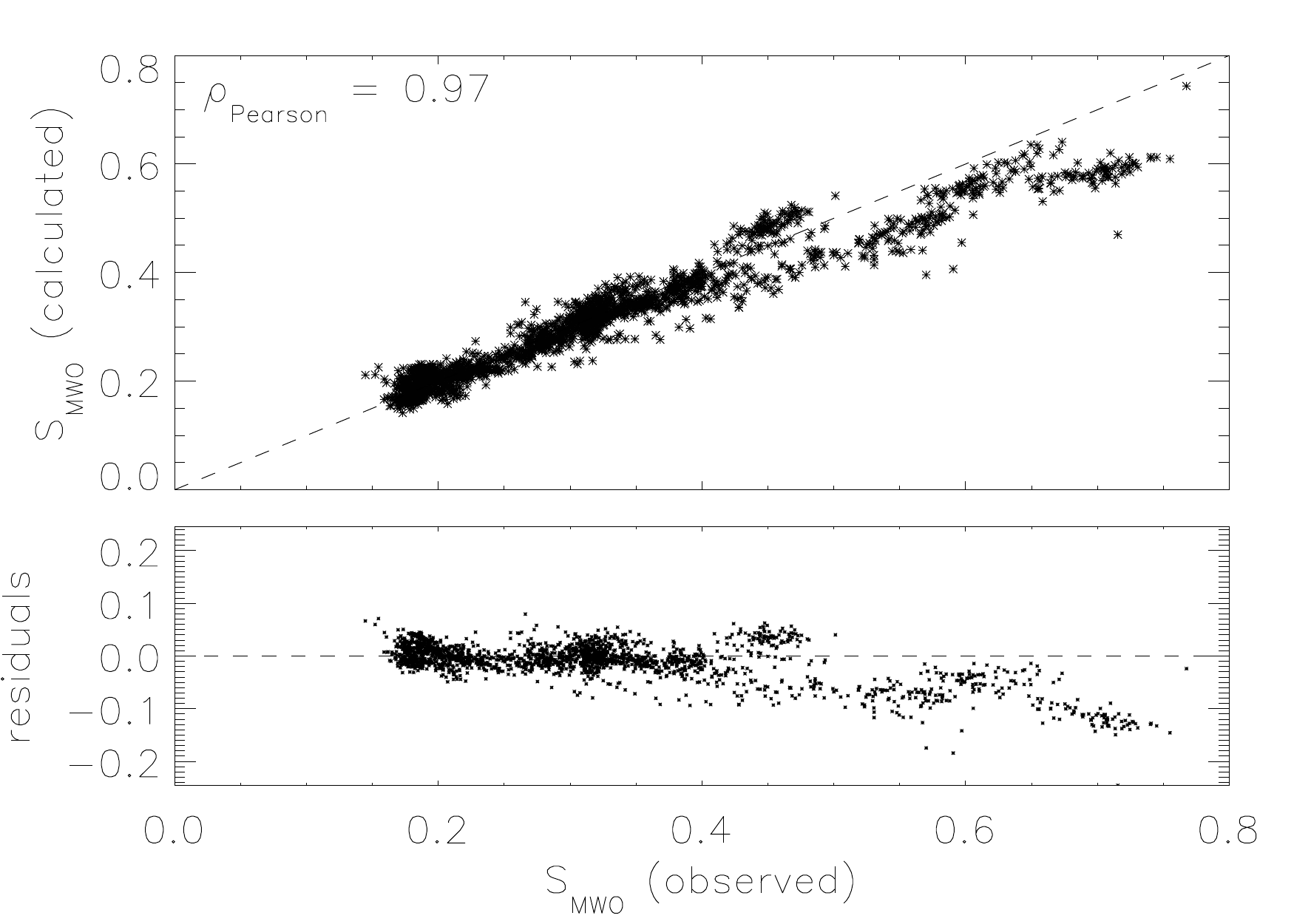}}
\caption{Comparison of the measured value for $S_\mathrm{MWO}$ to the one converted from the excess flux in the \ion{Ca}{II}~IRT lines, using Eq. \ref{eq:s_from_irt}. This plot includes data from 2076 observations of 76 stars. The dashed line corresponds to the identity relation \textit{(top)}, or the zero-level \textit{(bottom)}. The Pearson correlation coefficient of converted to observed values is 0.97, indicating a good conversion.}
\label{fig:Scompare}
\end{figure}

We have already shown the activity indices $S_\mathrm{MWO}$ and $R_\mathrm{HK}'$, which are both widely used. In \citet{Mittag2013}, the authors define an additional index that does not include basal flux contributions:

\begin{equation}
\label{eq:rhkpdef}
 R_\mathrm{HK}^{+}  =  \frac{F_\mathrm{HK}-F_\mathrm{HK,phot}-F_\mathrm{HK,basal}}{\sigma T_\mathrm{eff}^4} = \frac{F_\mathrm{Exc,HK}}{\sigma T_\mathrm{eff}^4}, 
\end{equation}

with $F_\mathrm{HK,basal}$ as the basal chromospheric flux contribution.\\
Both $R_\mathrm{HK}'$ and $R_\mathrm{HK}^{+}$ show a strong correlation ($\rho\geq0.9$) to our excess flux obtained here.\\
Since we have already provided relations to convert \ion{Ca}{II}~IRT measurements to $F_\mathrm{Exc,HK}$, converting them to $R_\mathrm{HK}^{+}$ is simply a matter of dividing by $\sigma T_\mathrm{eff}^4$. This parameter can
be estimated from $B-V$, so it could be included in the fit.\\
Here, we find the conversion from \ion{Ca}{II}~IRT-measurements to the indices in Eq. \ref{eq:rhkdef} and Eq. \ref{eq:rhkpdef}, calculated from $S_\mathrm{MWO}$ using the relation in \citet{Mittag2013}. This allows us to compare the equations to convert to $R_\mathrm{HK}'$ and $R_\mathrm{HK}^{+}$ in a more consistent fashion than if we used the measured value $F_\mathrm{Exc,HK}$, as we have not measured a $F_\mathrm{HK, chrom}$ that still includes a basal flux contribution. However, $F_\mathrm{Exc,HK} / \sigma T_\mathrm{eff}^4$ and $R_\mathrm{HK}^{+}$ are very close to identical, except for an offset already discussed in Sect. \ref{sec:fhkflux}. We find similar parameters for the formulae when using a value for $R_\mathrm{HK}^{+}$ determined using our measured $F_\mathrm{Exc,HK}$. Applying the method described in Sect. \ref{sec:conversion} yields:
\begin{eqnarray}
\label{eq:rhk_from_irt}
 R_\mathrm{HK}' & = & m \cdot F_\mathrm{Exc,IRT}  + b,\mathrm{~with}\\ 
\log{m} &=& -10.014 - 1.815 \cdot (B-V) + 1.501 \cdot (B-V)^{2} \nonumber \\
b &=& -0.277\cdot10^{-4} + 1.069\cdot10^{-4} (B-V) \nonumber -\\
  & & 0.586\cdot10^{-4} (B-V)^{2} , \nonumber
\end{eqnarray}
\begin{eqnarray}
\label{eq:rhkplus_from_irt}
 R_\mathrm{HK}^{+} & = & m \cdot F_\mathrm{Exc,IRT} + b,\mathrm{~with}\\
\log{m} &=& -10.257 - 1.127 \cdot (B-V) + 1.033 \cdot (B-V)^{2}\nonumber \\
b &=& -0.459\cdot10^{-4} + 1.334\cdot10^{-4} (B-V) - \nonumber\\
 & & 0.753\cdot10^{-4} \cdot (B-V)^{2} , \nonumber
\end{eqnarray}
with $F_\mathrm{Exc,IRT}$ in erg\,s$^{-1}$\,cm$^{-2}$. The error is again estimated from residuals and is roughly $5.5\cdot10^{-6}$ for both $R_\mathrm{HK}'$ and $R_\mathrm{HK}^{+}$, corresponding to an average error of $\sim10\%$. The relations for slope and intercept are very similar in shape, and in case of the slope $m$ also the actual function values. However, the intercepts $b$ shows a clear offset that stems from the small chromospheric basal flux correction that forms the difference in the two indices. For the determination of this conversion, we have removed 198 datapoints with very different S/N in the red and blue channels. The Pearson correlation coefficient between the converted and observed values is $0.97$.
Figure \ref{fig:RHKcompare} compares the converted to the measured values.\\
To convert \ion{Ca}{II}~IRT measurements into $S_\mathrm{MWO}$, the following relation can be used:
\begin{eqnarray}
\label{eq:s_from_irt}
S_\mathrm{MWO} & = & m \cdot F_\mathrm{Exc,IRT} + b,\mathrm{~with}\\ 
\log{m} &=& -6.500 - 2.165 \cdot (B-V) + 2.264 \cdot (B-V)^{2}\nonumber \\
b &=& 0.044 + 0.202 \cdot (B-V) - 0.013 \cdot (B-V)^{2} , \nonumber
\end{eqnarray}
where $F_\mathrm{Exc,IRT}$ has to be entered in erg\,s$^{-1}$\,cm$^{-2}$. We estimate the errors from the residuals to be $0.03$, a relative error of about 6\,\%. We find a Pearson correlation coefficient of $\rho_\mathrm{Pearson}=0.97$. The converted values are compared to the measured ones in Fig. \ref{fig:Scompare}.

\section{Conclusions}
We have analyzed more than two thousand spectra of almost a hundred main-sequence stars of type F, G and K obtained by the TIGRE telescope, which simultaneously records the spectral range of the \ion{Ca}{II} H \& K-lines, as well as H$\alpha$ and the \ion{Ca}{II} IRT. By carefully selecting an inactive comparison star of similar spectral type as the target star and artificially broadening the comparison star's spectrum to the target star's rotational velocity, we are able to derive a purely-activity related excess flux in the center of these chromospheric lines without any photospheric or basal flux contributions. This excess flux is obtained both in terms of fraction of the continuum and in physical units (i.e., as a flux in erg\,s$^{-1}$\,cm$^{-2}$), and it is free from any scatter from temporal variations.\\
This large sample of data shows that the excess flux in these lines are well correlated, with the Spearman correlation coefficients exceeding $\rho=0.9$ for the \ion{Ca}{II} lines, and $\rho\approx0.8$ for the correlation of H$\alpha$ and the \ion{Ca}{II} lines. Due to this strong correlation, it is possible to convert the observed excess flux of the \ion{Ca}{II}~IRT lines into the corresponding excess flux of the other lines, or into activity indices derived from them, despite the lower excess flux and subsquently lower sensitivity.\\
We provide such conversion relations and the errors on them, estimated directly from the residuals. The relations have been obtained by fitting the relations individually for stars with similar $B-V$, in order to remove any sampling bias. The given relations are valid for stars with $0.5 \lesssim B-V \lesssim 1.0$; they can be used to indirectly obtain values for activity indicators such as $R_\mathrm{HK}'$ or $S_\mathrm{MWO}$ from infrared spectra, for example, those expected from the GAIA mission. This makes it possible to compare these new measurements with the large amount of archival data available for these activity indices, which in turn allows new studies in the temporal behavior of these indices. To obtain the excess flux without an available comparison spectrum, we give empirically derived relations to estimate the photospheric and basal flux in the \ion{Ca}{II} IRT lines in Table \ref{tab:usedinactive}. Subtracting the value calculated from these relations from the flux measured in the \ion{Ca}{II}
 IRT 
lines of the active star in question returns the excess flux value, which can subsequently be converted into other quantities, if so desired.\\
We hope to increase the $B-V$ validity range of the relations when more M-dwarfs have been observed by TIGRE, assuming that the strong correlations between the lines still hold for later types. Spectra taken by CARMENES could then also be used to obtain a value of the activity indices derived from the \ion{Ca}{II} H \& K-lines, despite these lines falling outside the spectral range of CARMENES.

\begin{acknowledgements}
      Part of this work was supported by the
      \emph{Deut\-sche For\-schungs\-ge\-mein\-schaft, DFG\/} project
      number RTG 1351.
      This research has made use of the SIMBAD database, operated at CDS, Strasbourg, France.
      This research has made use of NASA's Astrophysics Data System.
      We thank our anonymous referee for helpful suggestions that improved the quality of this paper.
\end{acknowledgements}

\bibliographystyle{aa} 
\bibliography{citations} 


%

\Online

\begin{appendix}
\section{Stellar parameters used}
Here, in Table \ref{tab:objparas} and Table \ref{tab:compparas} we list the stellar parameters that we used in this work, and give their source.

\begin{table*}
\centering
\caption{\label{tab:objparas} Stellar parameters for the stars investigated in this work. References are numbered in superscript and given below. Values missing were inferred from a fit to the values of other stars.}
\begin{tabular}{llllllllll}
\hline
\hline\\[-0.3cm]
\textbf{Name} & \textbf{\textit{B-V}} & \textbf{log~\textit{g}} & \textbf{[Fe/H]} & \textbf{$v\sin{i}$} & \textbf{Name} & \textbf{\textit{B-V}} & \textbf{log~\textit{g}} & \textbf{[Fe/H]} & \textbf{$v\sin{i}$}\\
\hline\\[-0.2cm]
HD\,88355 & 0.43$^{(1)}$ & - & 0.00$^{(1)}$ &  - &
HD\,75332 & 0.50$^{(1)}$ & 4.41$^{(1)}$ & 0.13$^{(1)}$ & 11.00$^{(11)}$\\
HD\,25457 & 0.50$^{(1)}$ & 4.30$^{(1)}$ & -0.03$^{(1)}$ & 20.24$^{(3)}$&
HD\,179949 & 0.50$^{(1)}$ & 4.44$^{(1)}$ & 0.20$^{(1)}$ & 6.40$^{(2)}$\\
HD\,35296 & 0.52$^{(1)}$ & 4.28$^{(1)}$ & -0.02$^{(1)}$ & 16.00$^{(2)}$&
HD\,19019 & 0.52$^{(1)}$ & 4.00$^{(1)}$ & -0.17$^{(1)}$ & 10.00$^{(9)}$\\
HD\,20367 & 0.52$^{(1)}$ & 4.46$^{(1)}$ & 0.13$^{(1)}$ &  -  &
HD\,137107 & 0.55$^{(1)}$ & 4.22$^{(1)}$ & -0.03$^{(1)}$ &  - \\
HD\,100180 & 0.57$^{(1)}$ & 4.25$^{(1)}$ & -0.06$^{(1)}$ & 3.59$^{(3)}$&
HD\,150706 & 0.57$^{(1)}$ & 4.47$^{(1)}$ & -0.03$^{(1)}$ & 10.00$^{(9)}$\\
HD\,154417 & 0.58$^{(1)}$ & 4.38$^{(1)}$ & -0.01$^{(1)}$ & 8.00$^{(3)}$ &
HD\,206860 & 0.58$^{(1)}$ & 4.49$^{(1)}$ & -0.08$^{(1)}$ & 12.81$^{(3)}$\\
HD\,209458 & 0.58$^{(1)}$ & 4.45$^{(1)}$ & 0.01$^{(1)}$ & 4.5$^{(10)}$&
HD\,70573 & 0.59$^{(1)}$ & 4.58$^{(1)}$ & -0.11$^{(1)}$ & 19.39\,$\pm$\,4.00$^{(9)}$\\
HD\,114710 & 0.59$^{(1)}$ & 4.43$^{(1)}$ & 0.07$^{(1)}$ & 4.72$^{(3)}$&
HD\,115383 & 0.59$^{(1)}$ & 4.25$^{(1)}$ & 0.13$^{(1)}$ & 7.20~$\pm$~1.10$^{(7)}$ \\
HD\,129333 & 0.59$^{(1)}$ & 4.47$^{(1)}$ & 0.16$^{(1)}$ & 22.01~$\pm$~3.95$^{(9)}$ &
HD\,26913 & 0.60$^{(1)}$ & 4.49$^{(1)}$ & -0.02$^{(1)}$ & 1.83$^{(5)}$\\
HD\,39587 & 0.60$^{(1)}$ & 4.45$^{(1)}$ & -0.03$^{(1)}$ & 10.79$^{(3)}$ &
HD\,97334 & 0.61$^{(1)}$ & 4.35$^{(1)}$ & 0.06$^{(1)}$ & 7.74$^{(3)}$\\ 
HD\,75767 & 0.61$^{(1)}$ & 4.33$^{(1)}$ & -0.04$^{(1)}$ & 4.00$^{(4)}$ &
HD\,165401 & 0.61$^{(1)}$ & 4.39$^{(1)}$ & -0.41$^{(1)}$ & 13.90$^{(3)}$\\
HD\,190406 & 0.61$^{(1)}$ & 4.39$^{(1)}$ & 0.04$^{(1)}$ & 8.27$^{(3)}$&
HD\,25680 & 0.62$^{(1)}$ & 4.52$^{(1)}$ & 0.05$^{(1)}$ & 3.20$^{(5)}$\\
HD\,72905 & 0.62$^{(1)}$ & 4.53$^{(1)}$ & -0.08$^{(1)}$ & 11.21$^{(3)}$& 
HD\,197076 & 0.62$^{(1)}$ & 4.41$^{(1)}$ & -0.11$^{(1)}$ & 10.22$^{(9)}$\\
HD\,126053 & 0.63$^{(1)}$ & 4.43$^{(1)}$ & -0.38$^{(1)}$ & 3.08$^{(3)}$&
HD\,181321 & 0.63$^{(1)}$ & 4.42$^{(1)}$ & -0.01$^{(1)}$ & 13.00$^{(8)}$\\ 
HD\,30495 & 0.64$^{(1)}$ & 4.49$^{(1)}$ & -0.01$^{(1)}$ & 3.57$^{(5)}$&
HD\,38858 & 0.64$^{(1)}$ & 4.48$^{(1)}$ & -0.22$^{(1)}$ & 2.61$^{(3)}$\\
HD\,71148 & 0.64$^{(1)}$ & 4.36$^{(1)}$ & -0.00$^{(1)}$ & 12.37$^{(3)}$&
HD\,146233 & 0.65$^{(1)}$ & 4.42$^{(1)}$ & 0.03$^{(1)}$ & 4.07$^{(3)}$\\
HD\,140538 & 0.65$^{(1)}$ & 4.47$^{(1)}$ & 0.05$^{(1)}$ & 11.01$^{(3)}$ &
HD\,159222 & 0.65$^{(1)}$ & 4.34$^{(1)}$ & 0.10$^{(1)}$ & 3.01$^{(3)}$\\
HD\,190771 & 0.65$^{(1)}$ & 4.41$^{(1)}$ & 0.14$^{(1)}$ & 4.20$^{(5)}$&
HD\,20619 & 0.66$^{(1)}$ & 4.42$^{(1)}$ & -0.24$^{(1)}$ & 3.20$^{(3)}$\\
HD\,28099 & 0.66$^{(1)}$ & 4.43$^{(1)}$ & 0.13$^{(1)}$ & 3.54$^{(5)}$&
HD\,42618 & 0.66$^{(1)}$ & 4.46$^{(1)}$ & -0.11$^{(1)}$ & 4.40$^{(15)}$\\
HD\,20630 & 0.67$^{(1)}$ & 4.49$^{(1)}$ & 0.06$^{(1)}$ & 5.86$^{(3)}$&
HD\,43162 & 0.67$^{(1)}$ & 4.38$^{(1)}$ & -0.05$^{(1)}$ & 9.63$^{(3)}$\\
HD\,73350 & 0.67$^{(1)}$ & 4.46$^{(1)}$ & 0.11$^{(1)}$ & 4.00$^{(4)}$&
HD\,76151 & 0.67$^{(1)}$ & 4.46$^{(1)}$ & 0.08$^{(1)}$ & 3.58$^{(3)}$ \\
HD\,145825 & 0.67$^{(1)}$ & 4.46$^{(1)}$ & 0.03$^{(1)}$ & 3.10\,$\pm$\,1.20$^{(8)}$&
HD\,224930 & 0.67$^{(1)}$ & 4.41$^{(1)}$ & -0.77$^{(1)}$ & 4.07$^{(3)}$\\
HD\,42807 & 0.68$^{(1)}$ & 4.46$^{(1)}$ & -0.03$^{(1)}$ & 3.80$^{(5)}$ &
HD\,6582 & 0.69$^{(1)}$ & 4.50$^{(1)}$ & -0.80$^{(1)}$ & 4.17$^{(3)}$\\
HD\,10086 & 0.69$^{(1)}$ & 4.39$^{(1)}$ & 0.12$^{(1)}$ & 2.40$^{(4)}$ &
HD\,68017 & 0.69$^{(1)}$ & 4.46$^{(1)}$ & -0.44$^{(1)}$ & 1.49$^{(3)}$\\
HD\,111395 & 0.69$^{(1)}$ & 4.54$^{(1)}$ & 0.10$^{(1)}$ & 2.60$^{(5)}$&
HD\,101501 & 0.74$^{(1)}$ & 4.55$^{(1)}$ & -0.07$^{(1)}$ & 3.26$^{(3)}$\\
HD\,103095 & 0.75$^{(1)}$ & 4.63$^{(1)}$ & -1.34$^{(1)}$ & 9.28$^{(3)}$& 
HD\,184385 & 0.75$^{(1)}$ & 4.49$^{(1)}$ & 0.12$^{(1)}$ & 2.70$^{(5)}$\\
HD\,152391 & 0.76$^{(1)}$ & 4.47$^{(1)}$ & -0.05$^{(1)}$ & 3.06$^{(5)}$ &
HD\,82443 & 0.77$^{(1)}$ & 4.45$^{(1)}$ & -0.13$^{(1)}$ & 5.90$^{(5)}$\\
HD\,82885 & 0.77$^{(1)}$ & 4.49$^{(1)}$ & 0.32$^{(1)}$ & 7.22$^{(3)}$ &
HD\,131156A & 0.77$^{(1)}$ & 4.54$^{(1)}$ & -0.12$^{(1)}$ &  - \\
HD\,149661 & 0.78$^{(1)}$ & 4.50$^{(1)}$ & 0.03$^{(1)}$ & 1.63$^{(5)}$&
HD\,185144 & 0.78$^{(1)}$ & 4.49$^{(1)}$ & -0.22$^{(1)}$ & 6.79$^{(3)}$\\
HD\,100623 & 0.81$^{(1)}$ & 4.60$^{(1)}$ & -0.41$^{(1)}$ & 6.79$^{(3)}$&
HD\,165341A & 0.83$^{(1)}$ & 4.49$^{(1)}$ & -0.04$^{(1)}$ & 16.00$^{(11)}$\\
HD\,10476 & 0.84$^{(1)}$ & 4.45$^{(1)}$ & -0.05$^{(1)}$ & 1.20$^{(3)}$ &
HD\,115404 & 0.85$^{(1)}$ & 4.45$^{(1)}$ & -0.19$^{(1)}$ &  - \\
HD\,17925 & 0.86$^{(1)}$ & 4.52$^{(1)}$ & 0.07$^{(1)}$ & 4.80$^{(5)}$ &
HD\,97658 & 0.86$^{(1)}$ & 4.49$^{(1)}$ & -0.30$^{(1)}$ & 8.70$^{(5)}$\\
HD\,118972 & 0.86$^{(1)}$ & 4.36$^{(1)}$ & -0.02$^{(1)}$ & 4.10~$\pm$~1.20$^{(8)}$&
HD\,166620 & 0.87$^{(1)}$ & 4.47$^{(1)}$ & -0.17$^{(1)}$ & 4.82$^{(3)}$\\ 
HD\,75732 & 0.87$^{(1)}$ & 4.41$^{(1)}$ & 0.28$^{(1)}$ & 2.27$^{(3)}$&
HD\,22049 & 0.88$^{(1)}$ & 4.53$^{(1)}$ & -0.10$^{(1)}$ & 4.08$^{(3)}$ \\
HD\,37394 & 0.90$^{(1)}$ & 4.51$^{(1)}$ & 0.08$^{(1)}$ & 2.80$^{(5)}$&
HD\,4628 & 0.90$^{(1)}$ & 4.63$^{(1)}$ & -0.27$^{(1)}$ & 1.50$^{(5)}$ \\
HD\,145675 & 0.90$^{(1)}$ & 4.45$^{(1)}$ & 0.41$^{(1)}$ & 2.6$^{(10)}$&
HD\,22468 & 0.92$^{(1)}$ & - & - &  - \\
HD\,189733 & 0.93$^{(1)}$ & 4.49$^{(1)}$ & -0.02$^{(1)}$ & 2.30$^{(13)}$ &
HD\,5133 & 0.94$^{(1)}$ & 4.66$^{(1)}$ & -0.11$^{(1)}$ & 3.52$^{(3)}$\\
HD\,160346 & 0.96$^{(1)}$ & 4.46$^{(1)}$ & -0.03$^{(1)}$ & 3.37$^{(3)}$ &
HD\,16160 & 0.98$^{(1)}$ & 4.54$^{(1)}$ & -0.12$^{(1)}$ & 0.90$^{(5)}$\\
HD\,87883 & 0.99$^{(1)}$ & 4.47$^{(1)}$ & 0.05$^{(1)}$ & 1.20$^{(3)}$&
HD\,32147 & 1.06$^{(1)}$ & 4.41$^{(1)}$ & 0.26$^{(1)}$ & 5.18$^{(3)}$\\
HD\,131977 & 1.11$^{(1)}$ & 4.35$^{(1)}$ & -0.00$^{(1)}$ & 2.48$^{(5)}$&
HD\,190007 & 1.12$^{(1)}$ & 4.38$^{(1)}$ & 0.16$^{(1)}$ & 2.55$^{(5)}$\\
HD\,156026 & 1.16$^{(1)}$ & 4.60$^{(1)}$ & -0.20$^{(1)}$ & 4.40$^{(2)}$&
HD\,201091 & 1.18$^{(1)}$ & 4.70$^{(1)}$ & -0.38$^{(16)}$ & 4.72$^{(3)}$ \\
\hline
\end{tabular}
\tablebib{
  (1) \citet{Soubiran2010}; (2) \citet{Schroeder09}; (3) \citet{Martinezarnaiz10}; (4) \citet{Marsden14}; (5) \citet{Mishenina12}; (6) \citet{Bernacca70}; (7) \citet{AmmlervonEiff12}; (8) \citet{Torres06}; (9) \citet{White07}; (10) \citet{Glebocki05}; (11) \citet{Maldonado12}; (12) \citet{Takeda05}; (13) \citet{Torres12}; (14) \citet{Uesugi70}; (15) \citet{McCarthy14}; (16) \citet{Prugniel11}
  }
\end{table*}

\begin{table*}
\centering
\caption{\label{tab:compparas} Stellar parameters for the comparison stars used. References are numbered in superscript and given below.}
\begin{tabular}{llllllllll}
\hline
\hline\\[-0.3cm]
\textbf{Name} & \textbf{\textit{B-V}} & \textbf{log~\textit{g}} & \textbf{[Fe/H]} & \textbf{$v\sin{i}$} & \textbf{Name} & \textbf{\textit{B-V}} & \textbf{log~\textit{g}} & \textbf{[Fe/H]} & \textbf{$v\sin{i}$}\\
\hline\\[-0.2cm]
HD\,739 & 0.40$^{(1)}$ & 4.27$^{(1)}$ & -0.09$^{(1)}$ & 4.40$^{(9)}$ &
HD\,159332 & 0.45$^{(1)}$ & 3.85$^{(1)}$ & -0.23$^{(1)}$ & 5.00$^{(6)}$\\
HD\,216385 & 0.48$^{(1)}$ & 3.95$^{(1)}$ & -0.29$^{(1)}$ & 3.00$^{(4)}$&
HD\,45067 & 0.53$^{(1)}$ & 4.01$^{(1)}$ & -0.09$^{(1)}$ & 5.00$^{(6)}$\\
HD\,187691 & 0.56$^{(1)}$ & 4.26$^{(1)}$ & 0.10$^{(1)}$ & 3.00$^{(4)}$&
HD\,100180 & 0.57$^{(1)}$ & 4.25$^{(1)}$ & -0.06$^{(1)}$ & 3.59$^{(1)}$\\
HD\,124570 & 0.58$^{(1)}$ & 4.05$^{(1)}$ & 0.08$^{(1)}$ & 3.00$^{(4)}$&
HD\,19373 & 0.59$^{(1)}$ & 4.21$^{(1)}$ & 0.08$^{(1)}$ & 3.15$^{(1)}$\\
HD\,168009 & 0.60$^{(1)}$ & 4.23$^{(1)}$ & -0.01$^{(1)}$ & 3.00$^{(4)}$&
HD\,10307 & 0.62$^{(1)}$ & 4.32$^{(1)}$ & 0.03$^{(1)}$ & 4.07$^{(1)}$\\
HD\,34411 & 0.62$^{(1)}$ & 4.22$^{(1)}$ & 0.08$^{(1)}$ & 3.15$^{(1)}$&
HD\,95128 & 0.62$^{(1)}$ & 4.30$^{(1)}$ & 0.01$^{(1)}$ & 3.15$^{(1)}$\\
HD\,157214 & 0.62$^{(1)}$ & 4.31$^{(1)}$ & -0.40$^{(1)}$ & 3.15$^{(1)}$&
HD\,126053 & 0.63$^{(1)}$ & 4.43$^{(1)}$ & -0.38$^{(1)}$ & 3.08$^{(1)}$\\
HD\,38858 & 0.64$^{(1)}$ & 4.48$^{(1)}$ & -0.22$^{(1)}$ & 2.61$^{(1)}$&
HD\,146233 & 0.65$^{(1)}$ & 4.42$^{(1)}$ & 0.03$^{(1)}$ & 4.07$^{(1)}$\\
HD\,186427 & 0.65$^{(1)}$ & 4.32$^{(1)}$ & 0.07$^{(1)}$ & 2.18\,$\pm$\,0.50$^{(5)}$&
HD\,12846 & 0.66$^{(1)}$ & 4.38$^{(1)}$ & -0.26$^{(1)}$ & 2.20$^{(5)}$\\
HD\,42618 & 0.66$^{(1)}$ & 4.46$^{(1)}$ & -0.11$^{(1)}$ & 4.40$^{(8)}$ &
HD\,43587 & 0.67$^{(1)}$ & 4.29$^{(1)}$ & -0.04$^{(1)}$ & 2.98$^{(1)}$\\
HD\,3795 & 0.70$^{(1)}$ & 3.91$^{(1)}$ & -0.63$^{(1)}$ & 1.70$^{(2)}$&
HD\,115617 & 0.70$^{(1)}$ & 4.39$^{(1)}$ & -0.01$^{(1)}$ & 3.90\,$\pm$\,0.90$^{(3)}$\\
HD\,178428 & 0.70$^{(1)}$ & 4.25$^{(1)}$ & 0.14$^{(1)}$ & 1.50$^{(7)}$&
HD\,117176 & 0.71$^{(1)}$ & 3.97$^{(1)}$ & -0.06$^{(1)}$ & 4.83$^{(1)}$ \\
HD\,10700 & 0.72$^{(1)}$ & 4.48$^{(1)}$ & -0.50$^{(1)}$ & 1.60$^{(2)}$&
HD\,26965 & 0.85$^{(1)}$ & 4.51$^{(1)}$ & -0.27$^{(1)}$ & 2.10$^{(2)}$\\
HD\,75732 & 0.87$^{(1)}$ & 4.41$^{(1)}$ & 0.28$^{(1)}$ & 2.27$^{(1)}$&
HD\,145675 & 0.90$^{(1)}$ & 4.45$^{(1)}$ & 0.41$^{(1)}$ & 2.10$^{(2)}$\\
\hline
\end{tabular}
\tablebib{
  (1) \citet{Martinezarnaiz10}; (2) \citet{Jenkins11}; (3) \citet{AmmlervonEiff12}; (4) \citet{Takeda05}; (5) \citet{Marsden14}; (6) \citet{Bernacca70}; (7) \citet{Mishenina12}; (8) \citet{McCarthy14}; (9) \citet{Schroeder09};
}
\end{table*}

\end{appendix}

\end{document}